\documentclass{optica-article}

\journal{opticajournal} 

\articletype{Research Article}

\usepackage{subcaption}

\usepackage{lineno}
\linenumbers 

\def\revone{black}
\newcommand{\revtwo}[1]{\textcolor{black}{#1}}

\def\firsteditc{black}

\begin{document}
\nolinenumbers

\title{Adaptive E2E Monitoring Path Selection with Deep Reinforcement Learning in Multi-domain Optical Networks}

\author{Soham Choudhury,\authormark{1} 
Martin Bojinov,\authormark{1}
Jason P. Jue,\authormark{2} and 
Genya Ishigaki\authormark{1,*}}

\address{\authormark{1}Department of Computer Science, San Jose State University, San Jose, California 95112, USA\\
\authormark{2}Department of Computer Science, The University of Texas at Dallas, Richardson, Texas 75080, USA}

\email{\authormark{*}genya.ishigaki@sjsu.edu} 


\begin{abstract*} 
Network slicing over multi-domain optical networks enables resource isolation for high-bandwidth, virtually-dedicated networks. However, guaranteeing end-to-end (E2E) performance remains challenging when domains withhold internal topology and performance metrics from external entities. Under this limited visibility, a control function (slice coordinator) needs to select E2E monitoring paths to detect and localize link failures. This paper formulates two complementary optimization problems, the initial and progressive monitoring path selection problems, which together realize closed-loop autonomous monitoring at the slice coordinator level. The initial problem maximizes failure coverage without prior information, while the progressive problem concentrates additional paths around suspected failure locations identified from prior monitoring. We propose a deep reinforcement learning (DRL) algorithm that adaptively selects an optimal set of monitoring paths for each phase. Experiments using the GNPy optical network simulator demonstrate that our approach outperforms baseline algorithms for both problems and achieves near-optimal localization in the initial selection, suggesting the feasibility of AI-driven closed-loop failure localization in future multi-domain optical network architectures.
\end{abstract*}

\section{Introduction}

Network slicing is a technique that allows the instantiation of multiple virtual networks on a shared physical network infrastructure. Each virtual network is tailored to meet specific requirements, such as bandwidth and latency for different applications. The isolation of virtual networks enables satisfying the stringent service requirements of recent network applications \cite{8685766}.
Network slicing in the optical layer is promising for providing flexible and high-bandwidth slices. By abstracting optical network resources, network slicing simplifies the resource provisioning process for network applications and supports dynamic reconfiguration, while optimizing resource utilization across the network \cite{PORTELA2020103428}. To guarantee end-to-end (E2E) performance of such network slices, E2E network slice coordinators need to handle resource coordination and performance monitoring across multiple optical domains.

In particular, the autonomy of the optical network domains poses unique challenges to slice coordinators. When the domains are owned and managed by different entities, namely domain orchestrators, the orchestrators tend not to share the internal domain information, such as routing logic, topology, and detailed performance metrics, with other domains. This is to avoid unnecessary exposure of sensitive data and internal domain structure for security and operational flexibility. In a sense, the abstraction of the optical connectivity within a domain as optical connection services makes slice deployment simpler, but at the same time, the black-box connectivity makes it complex to guarantee the performance \cite{Casellas:20}. 

The focus of this paper is the E2E performance monitoring at the optical network slice level and localization of failures under the limited visibility of underlying resources from the slice coordinator's perspective. While it can be assumed that each domain monitors performance metrics, such as quality of transmission, within its administrative region, the multi-domain scenario requires the slice coordinator, with a holistic E2E view, to also measure and guarantee the performance by itself. As the domain orchestrators are not necessarily cooperative for E2E slice coordination tasks, we assume a minimal cooperation level to serve as a baseline level of monitoring. The minimal information that a domain can provide is its abstract connectivity services between its border nodes. This allows the E2E slice coordinator to see how domains are connected with each other via inter-domain links, leaving intra-domain information hidden and secure.

Assuming domain autonomy, the related works \cite{xu_hierarchical_2023, chen_cooperative_2022} solve routing, resource allocation, and failure recovery in multi-domain optical networks based on collaborative learning agents. 
The work in \cite{arrigoni_failure_2021} proposes an approximation algorithm for the Boolean network tomography problem of localizing node failures based on E2E monitoring \cite{duffield_network_2006}, assuming full visibility of networks. 
Under a similar setting, the localization capability of probing paths between randomly selected endpoints is formally evaluated in \cite{ma_node_2014}. 
Studies on routing with end-to-end feedback \cite{zhu_learning_2018} demonstrate that the structure of the network determines how much information a controller can infer solely from E2E measurements.
Several tomography works examine how topology influences the ability to localize failures. Progressive tomography \cite{arrigoni_failure_2021}, hierarchical routing methods \cite{xu_hierarchical_2023}, and SRLG localization \cite{ahuja_srlg_2011} all show that the network topology determines how effectively monitoring paths can distinguish between different failure locations. Crossed probing strategies \cite{gao_low-cost_2023} further highlight the importance of path diversity and overlap in enabling accurate localization. Partial tomography methods \cite{he_distributed_2018} also demonstrate that the choice of observed paths is critical when only limited measurements are available. 

In contrast to these related works, we focus on the selection of E2E monitoring paths for localizing failures under strict visibility constraints. Unlike prior approaches that assume fully known probing structures, our method addresses the problem of optimizing monitoring strategies from a high-level slice coordinator's perspective. In particular, the objective of the problem is to identify a set of E2E monitoring paths of maximal capability in the localization of single-link failures, given a constraint on the number of monitoring paths. 

This paper addresses two separate optimization problems: initial and progressive monitoring path selection problems, which collectively realize \textit{closed-loop autonomous monitoring} by a slice coordinator management function. The initial monitoring path selection problem is to decide a set of monitoring paths, without any prior information, to detect and localize more failures, while in the progressive monitoring path selection problem, the coordinator is required to add more monitoring paths to improve the localization capability, given the monitoring results from the initially deployed monitors. Although these two problems are the same in their output behaviors of selecting paths, their objectives are totally competing. The former needs to cover more links to catch a signal of \textit{any} failure in the network, while the latter is required to concentrate new monitoring paths around the suspected area to localize a \textit{specific} failure that was detected by the initial monitoring paths. 
 
Compared to node failure localization in related works, link failure localization is more complex due to the lack of overlap among selected monitoring paths. We devise a deep reinforcement learning (DRL) algorithm to learn the historical effectiveness of monitoring and select an optimal set of paths. In our experiments, the optical layer is simulated by a practical optical network simulator (GNPy), while the slice coordinator is implemented as a software module connected to the optical network through the north-bound API. Our experimental results demonstrate that our approach outperforms baseline algorithms in both the initial and progressive monitoring path selection problems and achieves near-optimal localization in the initial selection problem. Beyond the concrete results, our evaluation implies the possibility of an AI-based automation of failure detection and localization in future network architectures, which can dynamically adapt monitoring to the network environment. \textcolor{\revone}{Our simulator code and key datasets are available on our GitHub repository \cite{adaptive-e2e-monitoring}.}

Our earlier work \cite{choudhury2025reinforcement} proposed a DRL solution to the initial monitoring path selection problem. This paper extends our earlier work by discussing the new progressive monitoring problem and its DRL solution, which enhances the closed-loop autonomy of the monitoring function.

\section{Related Works}


\subsection{Multi-Domain Architectures and Monitoring Frameworks}
The foundations of multi-domain optical networking are surveyed extensively in \cite{chamania_survey_2009}, which highlights the challenges of peering, provisioning, and trust between administrative domains. Additional background on abstraction and resource exposure across heterogeneous domains is provided in \cite{zong_virtual_2020}, where virtual network embedding (VNE) is examined as a means of supporting multi-domain service provisioning despite limited visibility into each domain’s internal topology. Several architectures introduce hierarchical or cooperative control planes to coordinate domain behavior. Intelligent PCE-based multi-domain architectures \cite{hutchison_implementation_2014} and hierarchical routing schemes \cite{liu_hierarchical_2006} demonstrate how limited domain abstractions can be combined to support end-to-end path computation. More recent works explore autonomous and knowledge-based service provisioning \cite{chen_knowledge-based_2018} and hierarchical reinforcement learning for joint resource allocation \cite{xu_hierarchical_2023}, indicating growing interest in AI-driven inter-domain orchestration.

Monitoring and telemetry systems have also been studied extensively. \cite{velasco_monitoring_2019} outlines MDA (Monitoring \& Data Analytics) frameworks for optical networks, detailing architectures for collecting intra-domain telemetry. Meanwhile, TAPI-based telemetry streaming \cite{karunakaran_tapi-based_2024} defines vendor-agnostic mechanisms for exposing device data to controllers. Cooperative fault-management frameworks \cite{chen_cooperative_2022} propose hybrid-learning schemes that assume some degree of shared measurements between domains. These works, however, rely on intra-domain or per-hop telemetry, whereas our system only consumes E2E path observations generated from the broker’s abstracted topology.

\subsection{End-to-End Failure Localization and Network Tomography}
Much of the classic work on multi-domain localization is grounded in network tomography, where failures are inferred from end-to-end measurements. Progressive tomography \cite{arrigoni_failure_2021} and binary-performance tomography \cite{duffield_network_2006} show that link or node failures can be inferred by probing selected paths. Node-localization tomography \cite{ma_node_2014} further demonstrates that appropriately chosen end-to-end measurements allow precise node-failure detection. Optimization-based formulations \cite{khair_optimization_2009} and SRLG-aware localization \cite{ahuja_srlg_2011} refine this idea by incorporating network structure, risk groups, and routing constraints. Weighted matrix methods \cite{jung_effective_2011} and multi-fault localization mechanisms \cite{hong_yan_fast_2010} extend these models to multi-domain environments by correlating sets of failure indicators, typically alarms, across paths. From a localization perspective, these alarms can be viewed equivalently to our signaling monitoring paths, as both represent binary indicators that a failure has affected a set of links. However, our monitoring paths are not fixed. They are selected and optimized by an RL agent.


\textcolor{\revone}{
A classical line of work studies probe- and monitor-selection for failure localization under identifiability guarantees. Ma et al. \cite{6848079} characterize monitor placement for maximal identifiability, and Bartolini et al. \cite{8057091} derive fundamental bounds on failure identifiability in Boolean network tomography, while Arrigoni et al. \cite{9875226} progressively select probing paths by expected information gain over a posterior distribution of failure hypotheses. These methods assume that monitoring paths can be formed freely over a fully known topology. Our setting instead imposes a piggyback constraint: monitoring is carried only on pre-established lightpaths, so the feasible measurement set is a fixed subset determined by the deployed optical services rather than all paths defined in a given topology. This constraint changes the underlying search space, which is strictly smaller and irregular than the space assumed in the existing greedy path selection methods. In this context, the existing approaches provide only loose bounds in our problem, and their selection criteria do not directly transfer without redefinition. For this reason, we propose our own RL approach to solve the monitoring path selection problems and benchmark against the problem-specific optimum computed under our piggyback constraint, which serves as a tighter and more meaningful reference.
}

More recent work has explored improving probing efficiency and operating under limited observability. Crossed probing path planning \cite{gao_low-cost_2023} reduces the number of required monitoring paths while maintaining localization capability through structured path overlap. Similarly, partial network tomography approaches \cite{he_distributed_2018} show that failures can still be detected even when only a subset of end-to-end measurements is available. 
Our problem follows these seminal works on theoretical localization problems, while exploring a learning-based solution, compared to the traditional combinatorial algorithms.

\textcolor{\revone}{Hu et al. \cite{10.1145/3773966.3777921} propose DeepNT, a path-centric graph neural network that infers unobserved path performance and network topology from a fixed set of end-to-end measurements. Whereas such methods focus on inferring performance from a given measurement set, our work addresses the complementary problem of actively selecting which paths to measure for failure localization under a probing budget.}

\textcolor{\revone}{
Two recent works apply reinforcement learning to measurement-path selection in network tomography. Feng et al. \cite{10.1109/TNET.2021.3118006} address bandwidth tomography to estimate path bandwidth from end-to-end measurement. They develop an RL method to construct measurement paths that yield the tightest bounds for unidentifiable links. The work in \cite{10298460} combines tomography with a multi-armed bandit formulation, selecting probing paths to gather delay information and identify the best-performing route. Both share our high-level idea of learning where to measure, but neither is directly comparable to our setting. The former assumes a static topology in an upper network layer, unlike our failing physical paths, while the latter's bandit objective drives exploration toward estimating an optimal route rather than achieving failure-localization coverage. Moreover, to the best of our knowledge, no prior work applies reinforcement learning to monitoring-path selection for binary failure localization, where the measurement objective is to discriminate the failed component rather than to estimate a continuous transmission quality metric.}

\subsection{Machine Learning and Data-Driven Fault Management}
Machine learning for optical networks has primarily focused on impairment estimation and fault classification. Hybrid-learning-assisted frameworks \cite{wang_hybrid-learning-assisted_2021} predict QoT or impairment levels based on device telemetry, while cooperative ML approaches \cite{chen_cooperative_2022} combine local domain models to classify failure types. Hierarchical RL work in multi-domain elastic networks \cite{xu_hierarchical_2023} applies reinforcement learning for routing and spectrum allocation, but not for monitoring. In contrast, our work applies RL to select monitoring paths themselves, a problem that is typically handled by heuristics or tomography-based combinatorial selection.

\subsection{Telemetry and Simulation for Localization}
Physical layer-aware simulation tools such as GNPy \cite{ferrari_gnpy_2020} model realistic OSNR/SNR behavior and are widely used for path feasibility and physical impairments. Recent demonstrations of optical network tomography on live, multi-domain infrastructure \cite{sasai_optical_2025} show growing interest in combining real-world telemetry with simulation-assisted inference. Our methodology leverages GNPy to simulate the physical-layer signal degradation on both monitoring trails and background traffic, enabling controlled soft-failure injection that is difficult to perform on production networks.



\section{The E2E Monitoring Path Selection Problem}

The scope of this paper is to localize single-link failures to their corresponding edge in a broker graph, using a limited number of monitoring paths. The effectiveness of a set of monitoring paths can be defined in two different ways, depending on their goals: coverage and targeting. \textit{Coverage} aims to distribute monitoring resources broadly across the network to maximize the likelihood of observing anomalous behavior on average, without prior assumptions about where faults may occur. In contrast, \textit{targeting} concentrates monitoring resources around network regions identified as suspicious by initial path monitoring, so that it can narrow down the failing candidates at the cost of reduced breadth. In this section, we formally define the system model, localization metrics, and two monitoring path selection problems, which correspond to the coverage and targeting concepts. 

\subsection{System Model}
A multi-domain optical network that hosts the E2E network slices can be modeled with a hierarchical architecture, where multiple domain operators and an E2E slice coordinator reside in the respective layers. We follow a standard abstraction used in related works, such as \cite{xu_hierarchical_2023}. The architecture consists of two network views: physical topology and broker graph topology. 

The physical topology is a representation of the entire infrastructure that includes all optical switch nodes and fiber links within and between domains. Each operator maintains a complete local view of its own infrastructure (nodes and links) but does not have access to the topology information of other domains. Therefore, in reality, there is no one who oversees the entire physical topology in a multi-domain optical network where domains are administered by different entities. 

\begin{figure}[t]
    \centering
    \includegraphics[width=0.7\columnwidth]{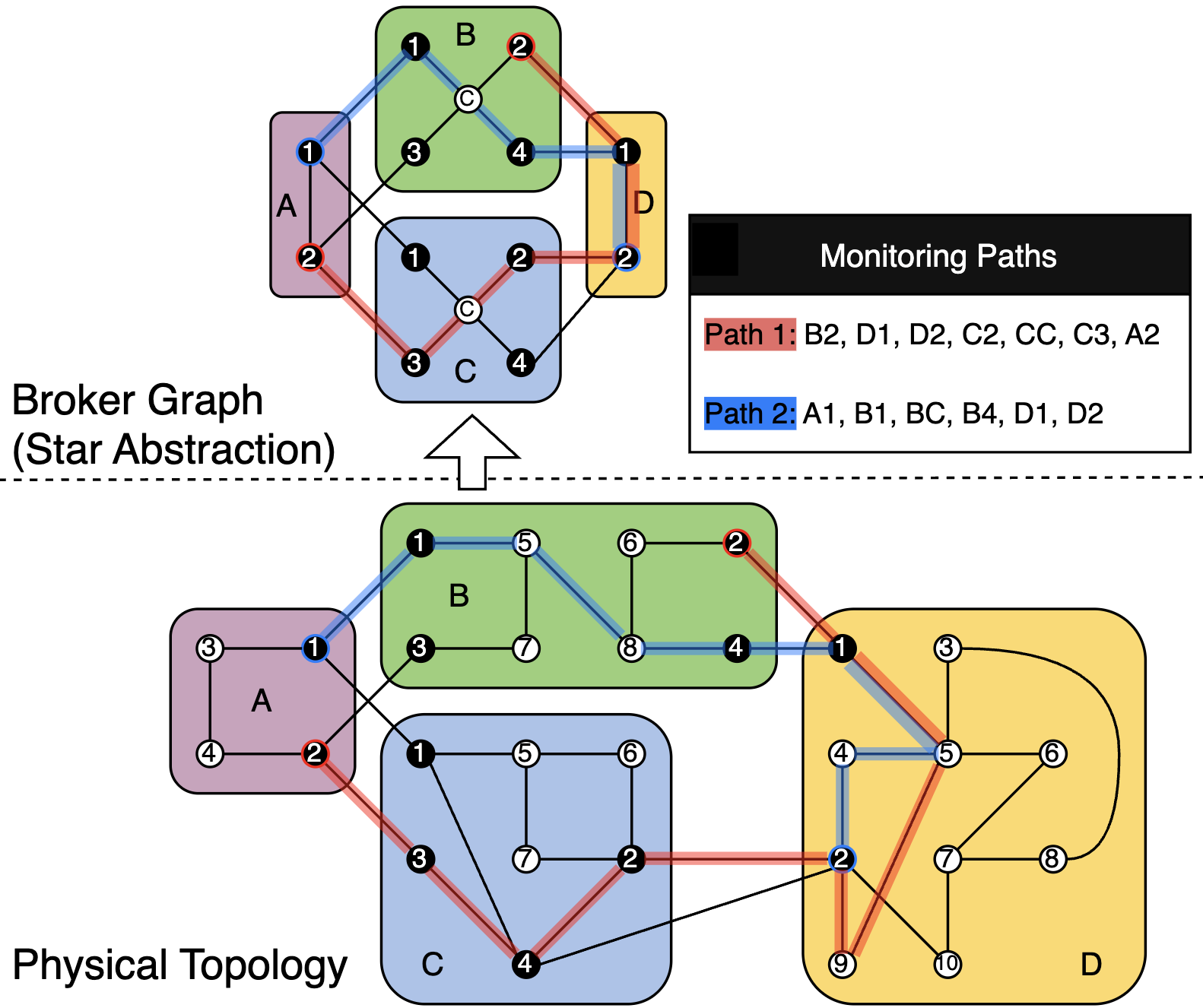}
    \caption{Topology Abstraction and a Broker Graph: Domains A-D are aggregated in the broker graph with the star graph abstraction.}
    \label{fig:abstraction}
\end{figure}

The broker graph topology is a network view provided to an E2E slice coordinator who communicates with domain operators and decides E2E slice routing and resource allocation. In this broker graph, each domain is aggregated to hide the internal details, as shown in Fig. \ref{fig:abstraction}. \textcolor{\firsteditc}{For the rest of the paper, a link refers to a physical fiber link, while an edge refers to the logical connection in the broker graph.} We assume the most strict scenario where a domain only reveals its \textit{border nodes} (ingress and egress routers) of the domain.

Given a set of border nodes, the E2E slice coordinator represents the logical connectivities among the border nodes by creating a star graph. A star abstraction represents the connectivities by edges between a central auxiliary node and all border nodes. Note that, in order to further reduce topological complexity, a star abstraction directly connects two border nodes without the auxiliary node when there are only two nodes in a domain.
Since the internal routing is not known, each edge in the two-hop path between border nodes in the star graph corresponds to a physical path within the domain. From the E2E slice coordinator's perspective, it uses the optical connectivity in the physical layer as a service to compose E2E network slices. 

\subsection{Localization Function}
To define an E2E monitoring problem for the E2E slice coordinator, we define the localization function that quantifies the ability of a set of paths to localize failures to the level of abstract edges in a broker graph. 
Each monitoring path $p$ corresponds to a lightpath in the broker graph $G$ and observes the E2E performance metric $M(p)$ of its corresponding monitored lightpath. We assume a realistic scenario to piggyback the monitoring signals onto the existing lightpaths. Hence, a monitoring path $p$ needs to be selected from all existing lightpaths $\mathcal{P}(G)$ already deployed in the network. 
\textcolor{\firsteditc}{In the hard failure model, a failure is defined based on a threshold $t$. When the performance metric of the monitored path drops below the threshold (i.e., $M(p) < t$), there should be at least one edge in the broker graph that is experiencing degradation on the lightpath.}

\def\mtrails{P(G)}
\def\ftrails{\rho}
\def\LNI{\text{LNI}}

Let $\mtrails \subseteq \mathcal{P}(G)$ denote a set of monitoring paths, and $\ftrails \subseteq \mtrails$ be a subset of the monitoring paths named \textit{signaling monitoring paths} that are experiencing a performance degradation, given a threshold value $t$:
\begin{align}
    \ftrails = \{p \in \mtrails \mid M(p) < t\}.
    \label{eq:signalingdef}
\end{align}
\textcolor{\revone}{For convenience, we use $\overline{\ftrails}$ to represent its complement set defined as $\overline{\ftrails} = P(G) \setminus \ftrails.$} 

We can identify a list $\tau(\ftrails)$ of edges that are potentially experiencing degradation (named \textit{potentially failing edges}) by taking the intersection of all the edges in the signaling monitoring paths $\textcolor{\revone}{p \in}\ftrails$, \textcolor{\revone}{minus the union of all edges in nonsignaling monitoring paths $p \notin \ftrails$} as follows.    
\begin{equation}
\tau(\ftrails) = \bigcap\limits_{p \in \ftrails} E(p) \textcolor{\revone}{ - \bigcup \limits_{p' \in \overline{\ftrails}} E(p')},
\label{eq:localize}
\end{equation}
where $E(p)$ is the set of edges in the broker graph on a given monitoring path $p$. \textcolor{\firsteditc}{Our objective is to identify an edge in the broker graph that is the root cause of the performance degradation of monitoring paths. Such an edge may represent multiple physical fiber links in the underlying physical topology. }

For a single-link failure scenario, we can perfectly localize the failing edge in the broker graph when the size of $\tau(\ftrails)$ is equal to 1. When $|\tau(\ftrails)| > 1$, the set of monitoring paths does not include enough paths to localize the failure to a specific edge in the broker graph.




\subsection{Initial Monitoring Path Selection}
The initial path selection problem is the optimization of monitoring path seeding, aiming at the coverage concept to maximize the monitoring capability on average across the network.
Mathematically, we can measure the coverage effectiveness of a set of monitoring paths $P(G)$ by intentionally failing each link in the graph. With each hypothetical failure, we can measure the size of $\tau(\ftrails)$ as monitoring performance and later sum them up to indicate the overall performance of the selected monitoring paths. We call this the Localization Narrowing Index (LNI). Since the set of \textcolor{\revone}{signaling} monitoring paths $\ftrails$ is a subset of $P(G)$, a choice of the monitoring paths changes the localization performance represented by the sum of $\tau(\ftrails)$. 
\begin{equation}
\LNI(P(G)) = \text{sum}_F( |\tau(\ftrails)|),
\label{eq:opt_coverage}
\end{equation}
where $F$ is a set of all possible single-link failure scenarios.

When every single-link failure can be localized, the size of $\tau(\ftrails)$ is always 1. In the single-link failure case, $F$ contains $|E|$ scenarios, where each corresponds to a failure of an edge in $E$. Therefore, the LNI of the set of monitoring paths with the perfect localization capability, $P^*(G)$, is always equal to the number of links in a given broker graph $G = (V,E)$: $\LNI(P^*(G)) = |E|$.


\def\allpsets{\mathcal{P}(G)}

Our E2E monitoring path selection problem is defined as the optimization of a set of monitoring paths to minimize the LNI defined in Eq. \eqref{eq:opt_coverage}.
\begin{align}
    \text{Minimize}_{P(G) \in \mathcal{P}(G)}\ \LNI(P(G)),
    \label{eq_obj}
\end{align}
where $\mathcal{P}(G)$ is a family of possible sets of monitoring paths in $G$.

As will be understood, a simple brute-force solution is not tractable to solve this problem in a large graph. The computation of the LNI requires iterating over all edges in the graph, and the minimization demands the LNI computation for all possible subsets of available paths, whose size could increase exponentially along with the size of the broker graph. 

As a nontrivial constraint, we define a cost constraint that bounds the number or length of monitoring paths that an E2E slice coordinator can select, with a constant budget $B$:
\begin{align}
    \sum_{p \in P(G)}c(p) \leq B.    
    \label{eq_const}
\end{align}
In this paper, we use a simple cost function $c(p) := 1\ (\forall p \in P(G))$ that limits the number of paths that an E2E slice coordinator can select to $K$ paths. It is easy to extend this constraint to a more general constraint by defining the actual cost to deploy a monitoring path $p$.





\subsection{Progressive Monitoring Path Selection}
Given the initial set of monitoring paths, a network operator observes the corresponding path telemetry information, such as OSNRs. When the budget $B$ is relatively tight compared to the network size, it is likely to have many potentially failing edges $\tau(\rho)$ after the initial monitoring. Therefore, we define the progressive monitoring path selection problem, where a network operator is allowed to add additional monitoring paths on top of the initial set of paths to further localize the failing edges. Here, the goal of the problem shifts away from coverage towards targeting, because the operator has some prior knowledge of network state based on the initial monitoring. The goal of the progressive path selection problem is to concentrate monitoring resources on the suspected regions of a network and to further reduce the number of candidate failing edges.

To formally describe the progressive addition of extra monitoring paths after the initial monitoring, let us define a logical time step $k = 0, ..., K$, where $k = 0$ denotes the initial path selection, and $K$ represents the total number of additional monitoring paths. We denote a set of the corresponding monitoring paths at each progressive time step as $P_k(G)$; i.e., a network operator selects one additional path $p^+_k\ (k = 1, ..., K)$ at each time step and builds an updated set of monitoring paths: 
\begin{align}
    P_k(G) = P_0(G) \cup \{p^+_1, ..., p^+_k\}\quad (k = 1, ..., K).
\end{align}
The initial monitoring provides the initial set of signaling monitoring paths: $\rho_0 = \{p \in P_0(G) \mid M(p) < t\}$ \textcolor{\revone}{if they exist. If $\{p \in P_0(G) \mid M(p) < t\}  = \emptyset$, $\rho_0 = P_0(G)$ to avoid the complete removal of the candidates.} The corresponding set of potentially failing edges is denoted as $\tau(\rho_0)$. The set of potentially failing edges at each time step is updated based on the signaling monitoring paths added:
\begin{align}
    \tau(\rho_k) =
    \begin{cases}
    \tau(\rho_{k-1}) \cap E(p^+_k)\quad &\text{if } M(p^+_k) < t\\
    \tau(\rho_{k-1}) \textcolor{\revone}{\setminus E(p^+_k)} &\text{otherwise}
    \end{cases}\quad (k = 1, ..., K).
\end{align}


Hence, the progressive path selection problem is the minimization problem of the size of a set of potentially failing edges at time step $K$ by selecting a set of additional monitoring paths $\{p^+_k\}$. This reflects the degree of ambiguity left in identifying the true failing link. Fewer remaining common edges indicate a more precise localization of the failure.
\begin{align}
    \text{Minimize }_{\{p^+_1, ..., p^+_K\}} |\tau(\rho_K)|,
\end{align}
given a set of initial monitoring paths $P_0(G)$ and the number of additional monitoring paths $K$.

\section{Reinforcement Learning-based E2E Monitoring Path Selection}
The computational complexity and failure dynamics over time motivate us to devise a learning solution for the monitoring path selection problem. In particular, we use deep reinforcement learning to learn an effective path selection strategy to localize failures. This learning module is implemented as a part of the E2E slice coordinator, whose network view is limited to a broker graph.

\subsection{Reinforcement Learning Agent}
We use Deep Q-Networks (DQN) with the target net \cite{mnih2013playingatarideepreinforcement} as the learning module in the following sections. DQN improves the estimate of the expected return of selecting a specific path $p$ at a network state $s$. Since the return is defined based on our objective in both problems, the agent eventually learns an optimal path selection. The estimates are represented as a parameterized Q function $Q(s,p;\theta)$ and updated through the optimization of the parameters $\theta$ as follows.
\begin{align}
    \theta \leftarrow \theta+\alpha \delta \nabla_\theta Q(s, p ; \theta),
\end{align}
where $\alpha$ is a learning rate, and $\delta$ is the temporal difference error defined with a target net $Q_T(s,p;\theta^{-})$ parameterized by $\theta^{-}$ and an immediate reward $r$ as 
\begin{align}
    \delta := r+\gamma \max _{p^{\prime}} Q_T\left(s^{\prime}, p^{\prime} ; \theta^{-}\right)-Q(s, p; \theta).
\end{align}

Our parameterized Q function is approximated by a neural network with one hidden layer consisting of 256 neurons and ReLU activation. A prioritized replay buffer of a capacity of 60,000 is used to improve the sample efficiency and the convergence. As its exploration method, we used a decaying epsilon greedy algorithm with the initial value of 1.0, with a decay rate of 0.0005. The learning rate $\alpha$ is set at 0.001. Also, we use a non-discounting environment with a discount factor $\gamma = 1.0$ to effectively propagate the reward signals in the sparse reward system defined below. The use of a target net stabilizes the learning by freezing the target value every 500 episodes.


\subsection{State and Action Spaces}
\subsubsection{Initial Monitoring Path Selection}
For the initial path selection, our agent observes a set of candidate monitoring paths as its state space. Each path is characterized by a list of nodes corresponding to optical switch nodes in the physical layer. Hence, each path is represented as a sequence of the integer node IDs. To limit the size of the candidate path set, we set an upper bound for the length of a path.
\textcolor{\revone}{In our implementation, we use a special data structure \textit{Repeated} from RLlib (Ray's RL library), which represents a variable-length list of a base space, up to a maximum length. Using the maximum bound of the path length, we can create a variable-length state space consisting of a list of paths (each path is a list of node IDs) for RL input.}

\textcolor{\firsteditc}{While the original initial monitoring path selection problem is defined as a one-shot optimization, we reformulate the problem as a progressive optimization problem, where the agent selects one path in each round until it hits the cost constraint. This means that the minimization in the objective function in Eq. \eqref{eq_obj} is sought by constructing a set of monitoring paths $P_0(G)$ by adding candidate paths one by one. After a path is selected, the path is removed from the candidate path set. The action space is discrete, where each action corresponds to selecting a candidate lightpath from a given set of existing lightpaths.}



\subsubsection{Progressive Monitoring Path Selection}

The state space for the progressive monitoring path selection problem consists of three components: (i) the set of currently selected monitoring paths encoded as sequences of node IDs (with padding to a fixed maximum length), (ii) associated physical-layer metrics $M(p)$ for each of the aforementioned paths, and (iii) the lengths of these paths. The lengths are used to indicate the padding indices to prevent an agent from learning misleading path features. 
\textcolor{\revone}{In the Gymnasium implementation, the observation space is defined as a dictionary of the three components that jointly encode the current set of selected monitoring paths. The first, chosen monitoring paths, is a matrix of shape ($M \times L$), where $M$ is the maximum number of monitoring paths, and $L$ is the maximum path length for padding. The second, OSNRs, records the OSNR measurements associated with each path, and the third, path lengths, stores the length of each path. Unused entries are padded with -1 to maintain a fixed-size representation suitable for the policy network.}
In our experiments, we use OSNRs as the physical-layer metrics. This representation allows the agent to reason jointly about topological features and signal quality. The action space remains the same from the initial monitoring path selection problem, where each action corresponds to the addition of one existing lightpath to the monitoring path set.


\subsection{Reward Function}
\subsubsection{Initial Monitoring Path Selection}
The definition of the LNI enforces us to define a sparse reward system for the initial monitoring path selection, where the reward at any state except for the terminal states is 0. This is because the potentially failing edges $\tau(\rho_t)$ can be identified only when all monitoring paths are selected as defined in Eq. \eqref{eq:localize}. At a terminal state in which the cost constraint is reached, we compute the immediate reward $r$ based on the set of selected monitoring paths $P(G)$ of a graph $G=(V,E)$ as follows.
\begin{equation}
r = \left(\dfrac{|E|^2}{\LNI(P(G))}\right)^s.
\label{eq:reward_func}
\end{equation}
The denominator, the LNI value, is upper bounded by the numerator, $|E|^2$; thus, the reward is between 1.0 and $|E|^s$ with any natural number $s$. 
Since the LNI is defined as the total of the sizes of the potentially failing edges for all failure scenarios, the tight upper bound of the LNI under single-link failures is $|E| \times |E|$, with the worst localization scenario of $\tau(\rho_t) = E$. The natural number $s$ is put in place to emphasize the difference between a better set of monitoring paths compared to a poorly performing one. Based on preliminary results, we used $s=3$ in our experiments. 


\subsubsection{Progressive Monitoring Path Selection}
The reward function for the progressive monitoring path selection is designed to encourage progressive reduction of the number of candidates of the failing edge. Specifically, the environment tracks a set of signaling monitoring paths $\rho_k$ that exhibit degraded OSNR ($M(p)$) below a given threshold $t$. When the agent selects a path that reduces the potentially failing edges $\tau(\rho_k)$ without eliminating all edges entirely, a small positive reward of 0.15 is assigned. Conversely, if the agent’s selection causes the size of $\tau(\rho_k)$ to collapse to zero, indicating loss of consistency among failure observations, a penalty with the value of -0.5 is applied. \textcolor{\revone}{Reward at each timestep was accumulated for all thresholds. In other words, training was not done individually for each threshold but done simultaneously based on a summed performance value across all predefined thresholds.}
This reward design encourages an agent to iteratively refine the candidate set while avoiding overly aggressive or inconsistent selections without clear \textit{targeting}.


\begin{figure}[t]
    \centering
    \includegraphics[width=0.8\columnwidth]{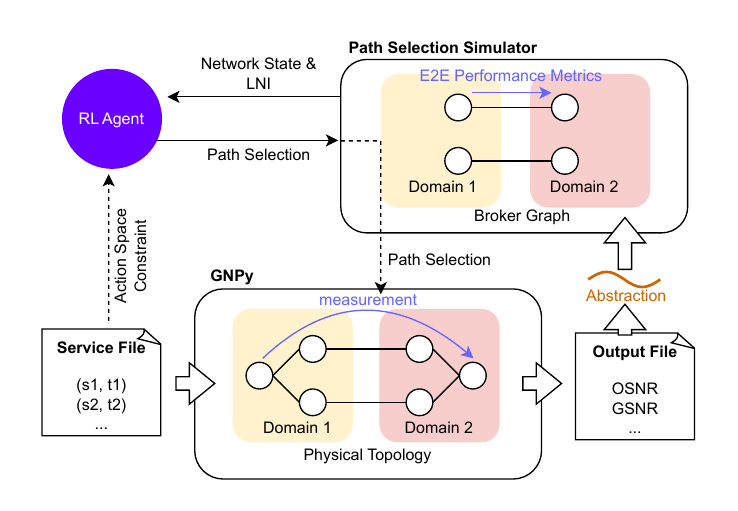}
    \caption{Simulator Setup: The physical layer is simulated based on GNPy, and a separate Python simulator that maintains the broker graph view is integrated. Our RL agent interfaces with the Python simulator to learn an optimal path selection.}
    \label{fig:overall_flow}
\end{figure}

\section{Evaluation Experiment}
To evaluate our RL-based approach in a realistic environment, we developed an optical network simulator using GNPy, an open-source Python library for modeling optical networks. The overall simulator setup is illustrated in Fig. \ref{fig:overall_flow}.

The physical graph topology is constructed as a simulated optical network in GNPy. GNPy is a Python library that supports simulation, evaluation, and optimization of optical network designs. In particular, it possesses the capability to report the generalized signal-to-noise ratio (GSNR) to within 1 dB of accuracy \cite{ferrari_gnpy_2020}. It enables network planners like our RL agent to assess feasibility and optimize performance by simulating the impact of nonlinear effects and noise in DWDM (Dense Wavelength Division Multiplexing) systems.

In GNPy, the input and output are given in the form of the \textit{service file} and \textit{output file}. A service file is used to define all lightpath requests to be propagated. For each request, the source and destination nodes, connection bandwidth, and spacing needs to be specified. Once these requests are fulfilled, GNPy returns an output that contains the information, such as the utilized network components, transmission quality metrics like OSNR, and lightpath information, such as its frequency and slot width.

The broker graph view is constructed by a separate simulator built on top of the GNPy simulator. Since the output file includes the complete view in the physical topology, we implemented an abstraction function that removes the internal topology and metrics within each domain and constructs a broker graph. Note that, in practice, this abstraction is performed by individual domain operators. We are simulating the independent entities using this functionality in our experiments.

Our RL agent interfaces with the wrapper simulator and only receives the network state and reward based on the E2E measurement metrics (e.g., OSNR) in the broker graph. An action of path selection is passed to the network simulator and reflected in the GNPy simulator so that the RL agent receives measurement results for the paths it selected. \textcolor{\revone}{In our experiment, separate RL agents were trained for each configuration of topology and candidate path set.}

\begin{table}[t]
    \centering
    \caption{Topology Properties.}
    \begin{tabular}{c|c|c|c|c}
        \hline
          & \# Domains & \# PHY nodes & \# Border nodes (by domain) & \# PHY links \\
         \hline
         Topology 1 & 3 & 16 & (1,2,3) & 24\\
         Topology 2 & 4 & 17 & (2,2,4,4) & 23\\
         Topology 3 & 6 & 46 & (4,4,4,4,4,10) & 81\\
         \hline
    \end{tabular}
    \label{tab:topos}
\end{table}

\subsection{GNPy Setup}    
In our experiments, three multi-domain optical network topologies are considered to test how well the agents perform across topologies of differing sizes and complexity. The topologies are constructed based on the works in \cite{liu_hierarchical_2006} and \cite{hutchison_implementation_2014}. The topological details are summarized in Table \ref{tab:topos}.


Each functional fiber has a length of 50 km and a loss coefficient of 0.2 dB/km. \textcolor{\revone}{Although the physical length of each fiber segment is fixed, the relevant abstraction is the broker graph, in which each node corresponds to a ROADM. Consequently, the effective length of an edge between two nodes is quantized to integer multiples of 50 km, and it yields a practical diversity in inter-node distances for our optimization problems.} Following the multiple OpenROADM design, we configure connections to be amplified immediately before reaching a ROADM and once again when it leaves the ROADM. The minimum slot width for a channel is 12.5 GHz; it can also only be increased by increments of 12.5 GHz. 
%
%
The bandwidth for all our requests is set to 50 Gigabits/second. For each request, the source and destination nodes are randomly selected among all ROADM nodes.

Our channel allocation algorithm is designed to first check the shortest available path between two transceivers for an available channel. If not, it will move onto the next available shortest path, and so on. The preference for shorter paths generally provides better performance of deployed lightpaths due to the reduction of travel distance.

\begin{table}[t]
    \centering
    \caption{E2E OSNR with Link Failures in Topology 2. \label{tab:top2_fiber_failure}}
    \begin{tabular}{c|p{2.2cm}|p{2.6cm}|p{2.6cm}|p{2.8cm}}
    \hline
    & Without Link Failure &Link Failure (Loss Coef.: 0.5 dB/km)&Link Failure (Loss Coef.: 1.0 dB/km)&Link Failure (Loss Coef.: 1.25 dB/km)\\
    \hline
    Path 1 & 19.55 & 19.55& 19.55&19.55\\
    Path 2 & 21.87 & 21.87& 21.87&21.87\\
    Path 3 & 20.1 & 20.1& 20.1& 20.1\\
    Path 4 & 19.55 & 18.88&10.4 &3.64\\
    \hline
    \end{tabular}
\end{table}


\subsection{GNPy Fiber Failure Simulation}
\subsubsection{Initial Monitoring Path Selection}
To simulate failure scenarios, we intentionally inject noise into a GNPy fiber cable. As the scope of this paper is single-link failure scenarios, noise is injected into only one fiber by degrading its loss coefficient parameter to 1.0 dB/km (hard failure). Relative to the setting, we used a failure threshold $t = 5.0$ in the experiments. 

Table \ref{tab:top2_fiber_failure} shows the effect of different levels of injected noise on OSNR in Topology 2. 
We see that the reported E2E OSNR for Path 4 suffers as the loss coefficient for the affected fiber is increased. In contrast, the other three paths, which do not include the affected fiber, did not experience any change in the reported E2E OSNR. 

\textcolor{\revone}{Because failures are inferred by comparing the reported E2E OSNR against the threshold $t$, measurement noise near the threshold can produce both false positives and false negatives, introducing the alarm ambiguity that the RL agent must contend with.}

\subsubsection{Progressive Monitoring Path Selection}


Similar to the setup for the initial monitoring path selection, we randomly select a fiber to inject the simulated noise of 0.8 dB/km (default is 0.2 dB/km) for the duration for an entire episode. All fibers are of length 50 km; however, an edge in our topologies may be made of more than one fiber, in which case adjacent fibers are conjoined by an amplifier. As we confirm the ability of our RL agent to detect the hard failure scenarios in the previous setup, we also simulated soft failure scenarios where the injected noise and the corresponding change in path OSNRs are mild, compared to the previous case, where we injected a trivial noise. Hence, we tried a few different threshold values $t = 16.5, 18.0$, and $20.0$ to define the signaling monitoring paths. We can analyze the impact of the threshold on the monitoring performances by comparing the results.




\subsection{Baseline Algorithms}
To evaluate our RL-based approach, we show the initial path selection results from three other baseline algorithms and the theoretical optimum, and the progressive path selection results from five baseline algorithms and the theoretical optimum. 

\begin{itemize}
    \item \textbf{Random}: Of the available paths between border nodes, select one path at random.
    \item \textbf{Longest Path}: Of the available paths between border nodes, select the option that covers the most edges. This is to maximize the number of edges a path covers.
    \item \textbf{Least Monitored}: For each candidate monitoring path, sum the number of pre-existing monitoring paths that cover each edge on the candidate path. Normalize the sum by dividing it by the length of the path, and select the path with the lowest score (the least monitored path). \textcolor{\revone}{The Least Monitored heuristic can be viewed as a coverage-inspired greedy method, since it incrementally selects paths that improve edge coverage.}
    \item \textbf{Theoretical Optimum} (Initial Monitoring Path Selection): When a set of given monitoring paths $\mathcal{P}(G)$ is sufficiently small, we can execute a brute-force search to find the optimum set of monitoring paths. 
    The red dotted lines in the following figures show the median of the theoretical optima. Note that we conducted 20 experiments and obtained the corresponding 20 different theoretical optimum sets of monitoring paths. Therefore, the plotted theoretical optimum could underperform another method due to the statistical processing. However, in our largest scenario with 16 monitoring paths in Topology 3, the brute force search was not able to find the optimum even after 72 hours of computation on a Linux server with the Intel i9 processor. \textcolor{\revone}{To obtain the optimal line for Topology 3 would require performing LNI calculation for 62,852,101,650 combinations for 20 time rounds.}
    \item \textcolor{\revone}{\textbf{Theoretical Optimum} (Progressive Monitoring Path Selection): We also show the theoretical optimum of the cardinality of a failing link candidate set for the progressive path selection problem. When a failed link is located in the area where monitoring path candidates are sparse, it is not possible to perfectly localize it, even when adding all the monitoring paths. We will show the minimum number of links that can be considered as potentially failing. Theoretically computing the minimum number is computationally cheap, because we can narrow down a set of monitoring paths we consider, given a ground-truth failing link. However, finding a set of monitoring paths to achieve the minimum number, which is our problem, is combinatorial and computationally difficult.}
    \item \textbf{Most Monitored} (Progressive Monitoring Path Selection \textit{Only}): Using the same metrics defined for the Least Monitored, we select an additional path for the targeting objective. After normalizing the sum of the number of monitoring paths that cover each edge, select the path with the highest score (the most monitored path) to concentrate the monitoring resources.
    \item \textbf{Minimal Overlap} (Progressive Monitoring Path Selection \textit{Only}): Randomly select a candidate monitoring path as long as it traverses one of the edges already covered by the set of pre-existing monitoring paths.
\end{itemize}

\section{Experiment Results}






\subsection{Initial Monitoring Path Selection Performance}
\begin{figure} 
    \centering
    \subfloat[4 paths.\label{fig:top1_star4}]{%
        \includegraphics[width=0.5\linewidth]{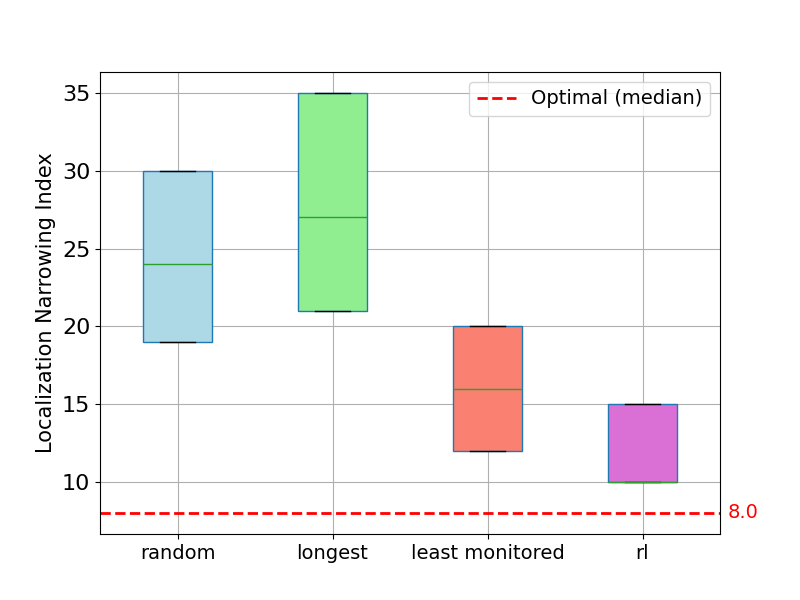}}
    \hfill
    \subfloat[8 paths.\label{fig:top1_star8}]{%
        \includegraphics[width=0.5\linewidth]{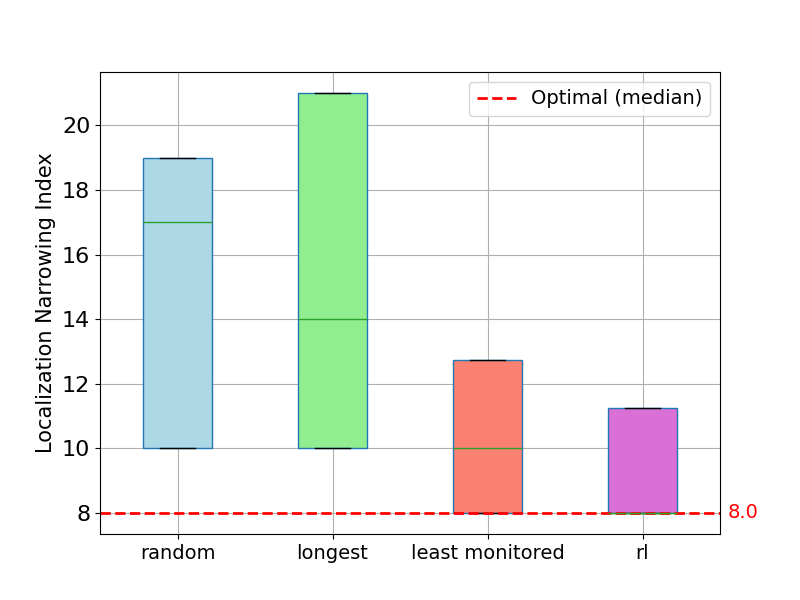}}
    \caption{Localization Narrowing Index on Topology 1.}
    \label{fig:top1 results} 
\end{figure}

\begin{figure} 
    \centering
    \subfloat[4 paths.\label{fig:top2_star4}]{%
        \includegraphics[width=0.5\linewidth]{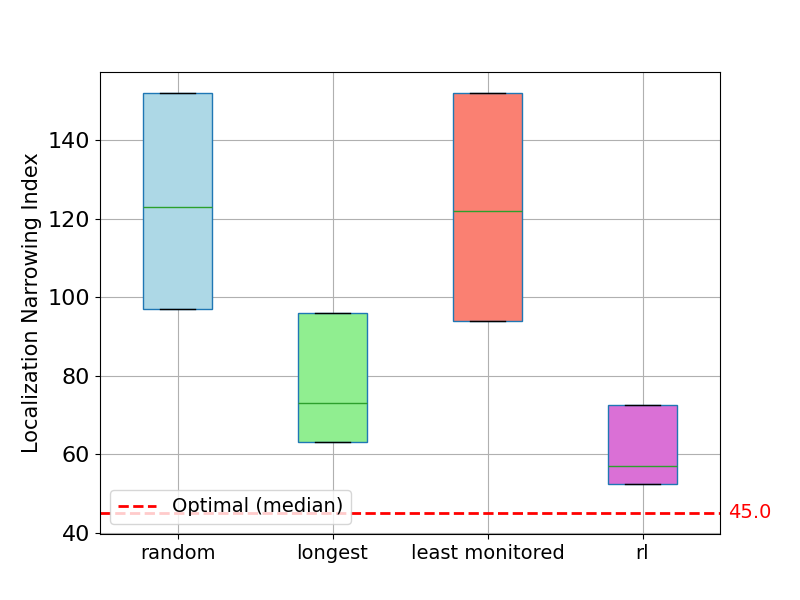}}
    \hfill
    \subfloat[8 paths.\label{fig:top2_star8}]{%
        \includegraphics[width=0.5\linewidth]{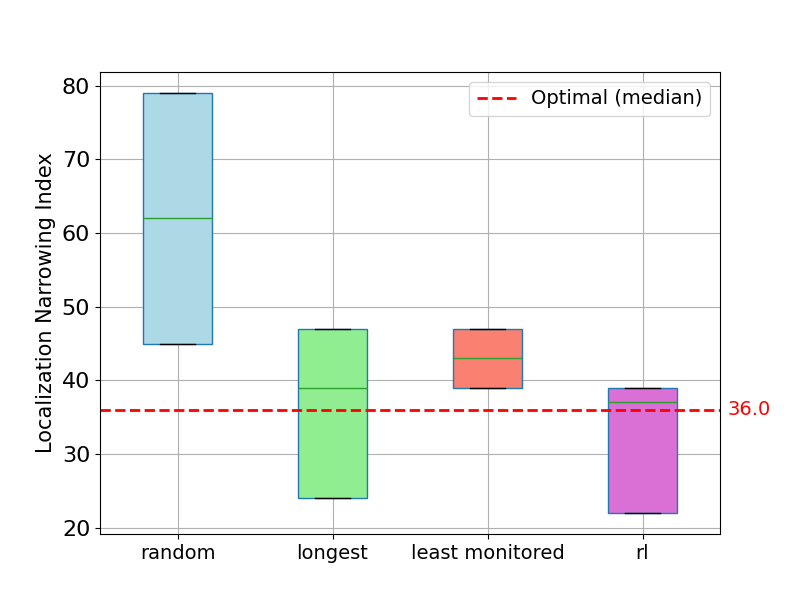}}
    \caption{Localization Narrowing Index on Topology 2.}
    \label{fig:top2 results} 
\end{figure}

\begin{figure} 
    \centering
    \subfloat[4 paths, 20 req.\label{fig:top3_star4}]{%
        \includegraphics[width=0.5\linewidth]{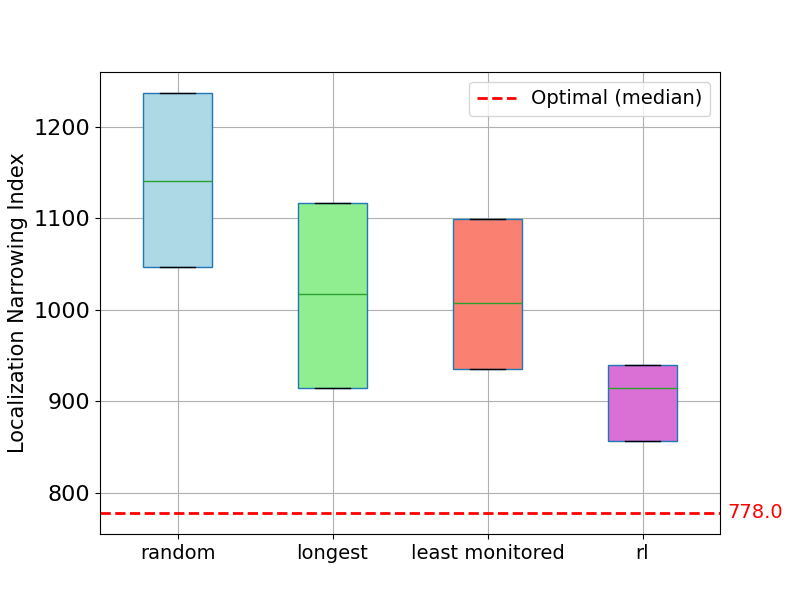}}
    \hfill
    \subfloat[8 paths, 20 req.\label{fig:top3_star8}]{%
        \includegraphics[width=0.5\linewidth]{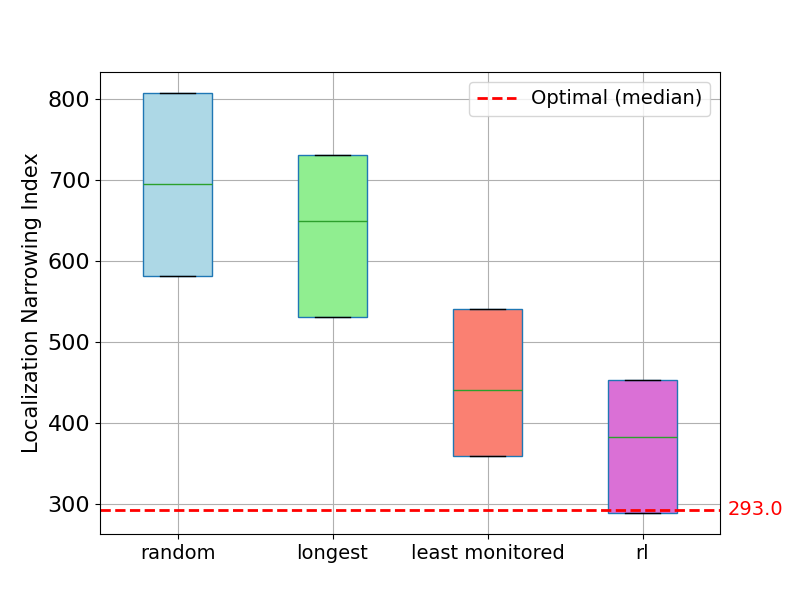}}
    \\
    \subfloat[16 paths, 40 req.\label{fig:top3_star16}]{%
        \includegraphics[width=0.45\linewidth]{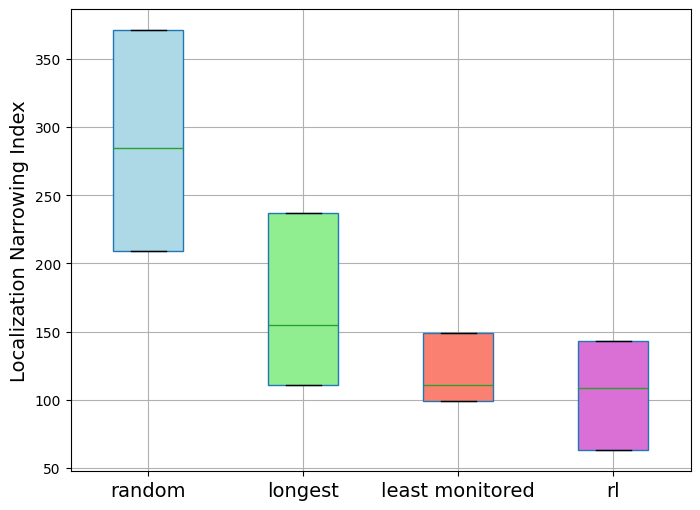}}
    \hfill
    \subfloat[Learning Curve (16 paths, 40 req.).\label{fig:learning}]{%
        \includegraphics[width=0.5\linewidth]{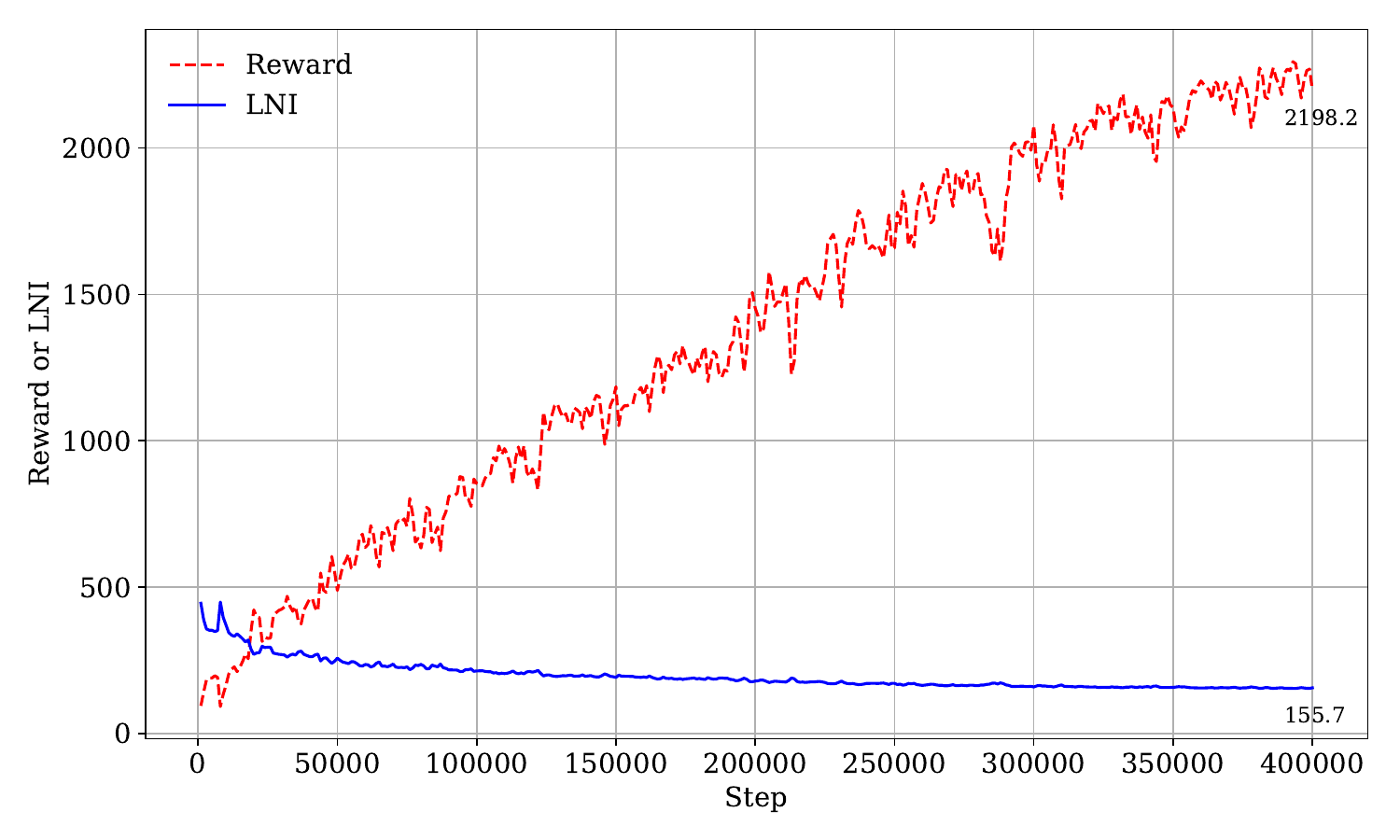}}
    \caption{Localization Narrowing Index on Topology 3.}
    \label{fig:top3 results} 
\end{figure}

We use 20 different configurations (service and output files) for each of our experiments. Each output file has information for 20 or 40 lightpaths, depending on the predefined cost constraint. All results shown below are collected and averaged for a hundred selection trials. The caption of each result in Figs. \ref{fig:top1 results}-\ref{fig:top3 results} shows the cost constraint bound ($B$ paths) defined in Eq. \eqref{eq_const}. Also, exemplary learning curves based on the LNI and reward function are included for Topology 3 in Fig. \ref{fig:top3 results}(d).

We observe that the LNI of our RL-based approach outperforms the heuristics across all experimental settings. Moreover, its results have a more compact spread due to the agent exploiting its best actions after policy convergence.

The broker graphs for Topology 1 are quite small with 8 edges. This incentivizes shorter monitoring paths with varying sets of edges. Longer paths are more likely to contain similar edges, hindering the ability to uniquely localize individual edges. This is reflected in the results \textcolor{\revone}{shown in Fig. \ref{fig:top1 results}}, where the RL agent and the least-monitored heuristic select shorter paths, scoring lower LNIs. The longest path heuristic suffers significantly in this environment. This is further compounded in the star abstraction, where there exist fewer unique paths. 

Topology 2 is more moderately sized, with its broker graph containing 18 edges. \textcolor{\revone}{Fig. \ref{fig:top2 results} shows the LNI obtained by all four methods.} The RL agent performs the best in \textcolor{\revone}{both} experiments, typically \textcolor{\revone}{followed by the longest path heuristic.} 

Topology 3 is a large, 6-domain environment, where its broker graph has 45 edges.
All experiments yield similar results, as illustrated in Fig. \ref{fig:top3 results}. Our RL performs the best, followed by least-monitored, longest path, and then random heuristics. Again, the theoretical optimum was not available due to the computational complexity with 16 monitoring paths in Topology 3. Note that 40 lightpaths were given to the RL agent in this specific case to provide more candidate path combinations.

\textcolor{\revone}{
To quantify how close each method's solution is to the theoretical optimum, we define the optimality gap as $\dfrac{(\text{LNI}_{\text{method}} - \text{OPT})}{\text{OPT}}$, where $\text{LNI}_{\text{method}}$ is the LNI value obtained by a given method, and $\text{OPT}$ is the optimum obtained via brute-force search. A smaller gap indicates a solution closer to optimal. We computed this gap for each of the 20 individual experiments.}
\textcolor{\revone}{RL accomplishes the lowest optimality gap across all topologies. In Topology 1, RL achieves a mean optimality gap of 0.2146 for 4 monitoring paths. With 8 monitoring paths, it averages a perfect 0 optimality gap. In Topology 2, RL reaches optimality gaps of 0.4595 and 0.0572 for 4 and 8 paths, respectively. Finally, in Topology 3, RL attains average optimality gaps of 0.1272 and 0.2403 for 4 and 8 paths, respectively.}

\textcolor{\revone}{Table \ref{tab:optgap} provides the detailed summary statistics of the optimality gap in Topology 3 with 8 candidate paths and 20 requests.
The RL approach achieves the lowest optimality gap across all statistics, with a mean of 0.2403 and a median of 0.2336, substantially outperforming the least-monitored (mean 0.5060), longest (mean 1.1694), and random (mean 1.3916) baselines. Moreover, RL exhibits the smallest standard deviation (0.1862) and the lowest maximum gap (0.6879), indicating not only superior average performance but also greater consistency across experiments.}

\begin{table}[t]
\centering
\caption{Comparison of Optimality Gap across Path Selection Methods in Topology 3 with 8 Candidate Paths and 20 Requests.}
\label{tab:optgap}
\begin{tabular}{r|ccccc}
\hline
Method & Mean & Median & Std & Min & Max \\
\hline
Random          & 1.3916 & 1.3278 & 0.5774 & 0.5010 & 3.0573 \\
Longest         & 1.1694 & 1.2514 & 0.6141 & 0.0948 & 2.5464 \\
Least Monitored & 0.5060 & 0.4444 & 0.2800 & 0.1438 & 1.2440 \\
RL              & \textbf{0.2403} & \textbf{0.2336} & \textbf{0.1862} & \textbf{0.0117} & \textbf{0.6879} \\
\hline
\end{tabular}
\end{table}


\begin{figure} 
    \centering
    \subfloat[The Size of Candidate Set. (True failing fiber: (dA-v2, dB-v3)). \label{fig:16_5_dA_v2_dB_v3_plot}]{%
        \includegraphics[width=0.48\linewidth]{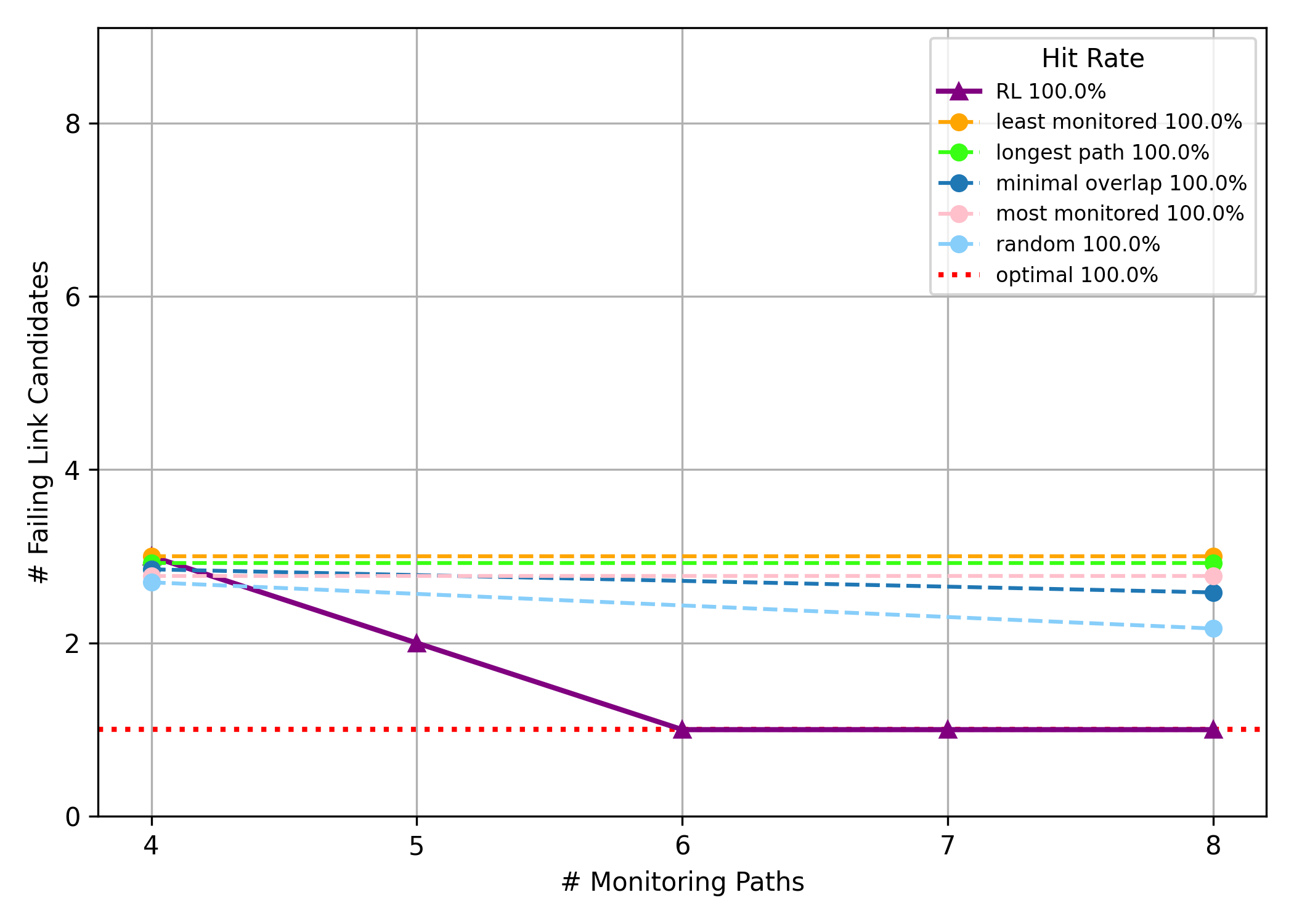}}
    \hfill
    \subfloat[The Efficiency Ratio. (True failing fiber: (dA-v2, dB-v3)). \label{fig:16_5_dA_v2_dB_v3_bar}]{%
        \includegraphics[width=0.48\linewidth]{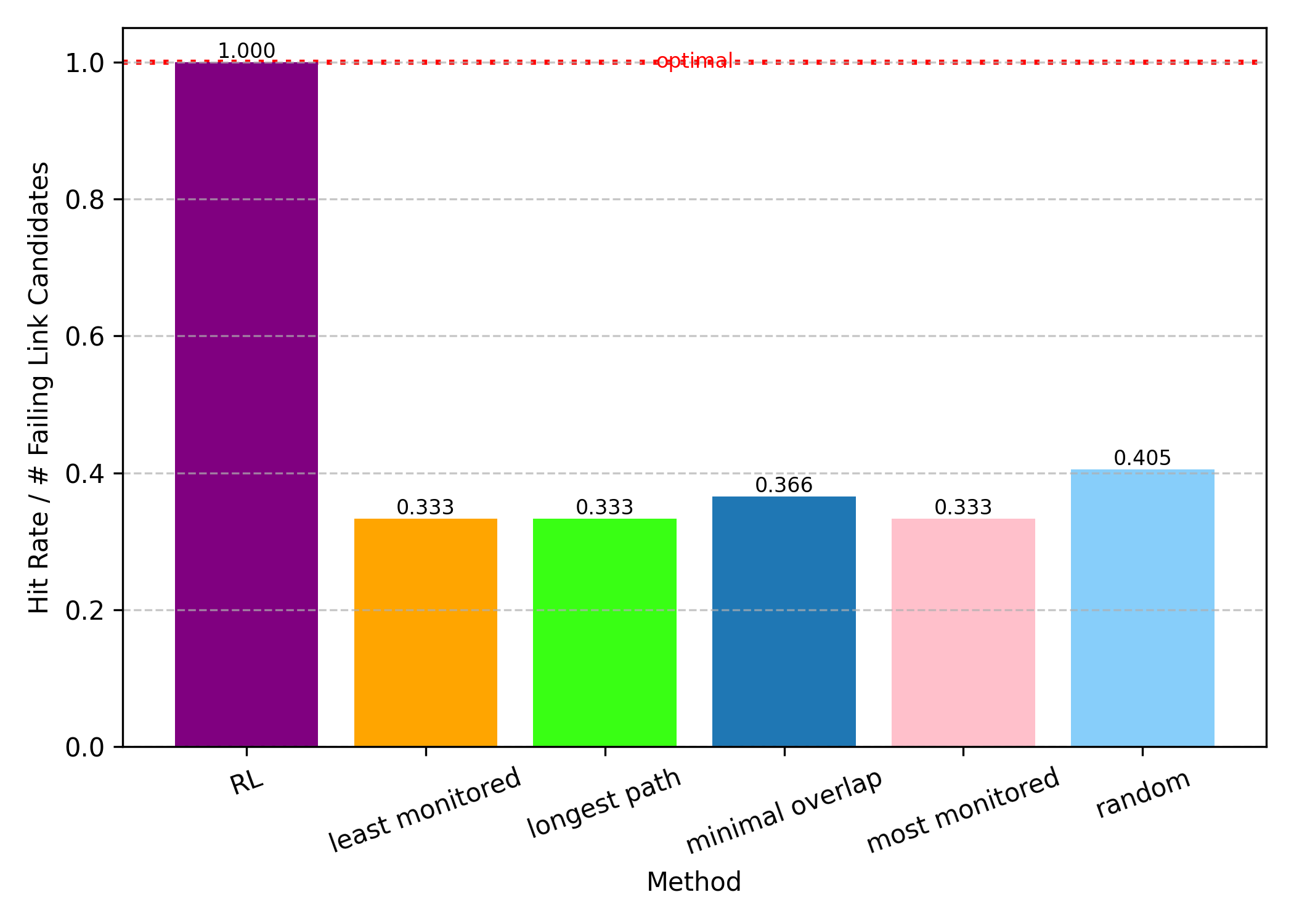}}
    \\
    \subfloat[The Size of Candidate Set. (True failing fiber: (dB-v4,dD-v1)). \label{fig:16_5_dB_v4_dD_v1_plot}]{%
        \includegraphics[width=0.48\linewidth]{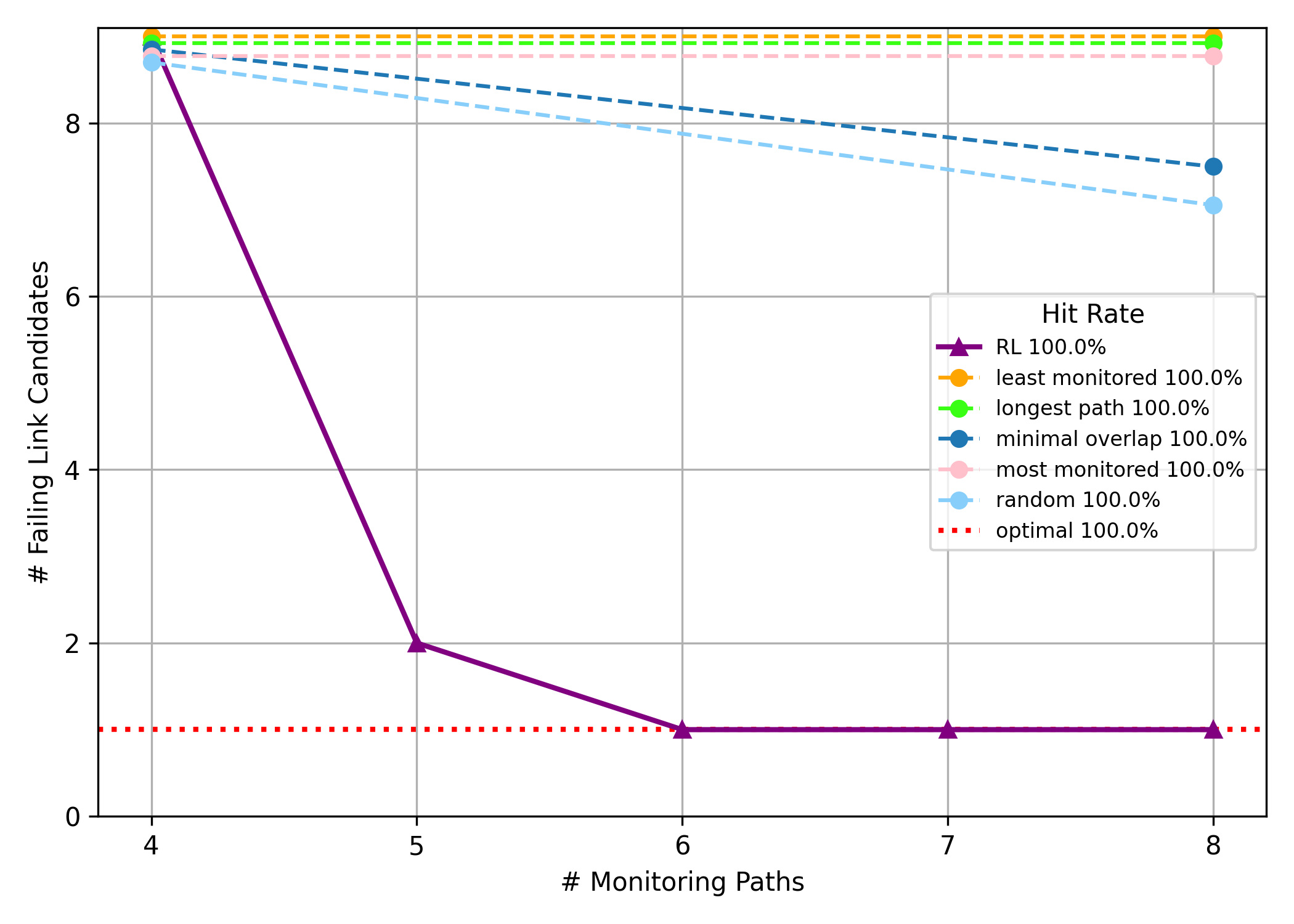}}
    \hfill
    \subfloat[The Efficiency Ratio. (True failing fiber: (dB-v4,dD-v1)).\label{fig:16_5_dB_v4_dD_v1_bar}]{%
        \includegraphics[width=0.48\linewidth]{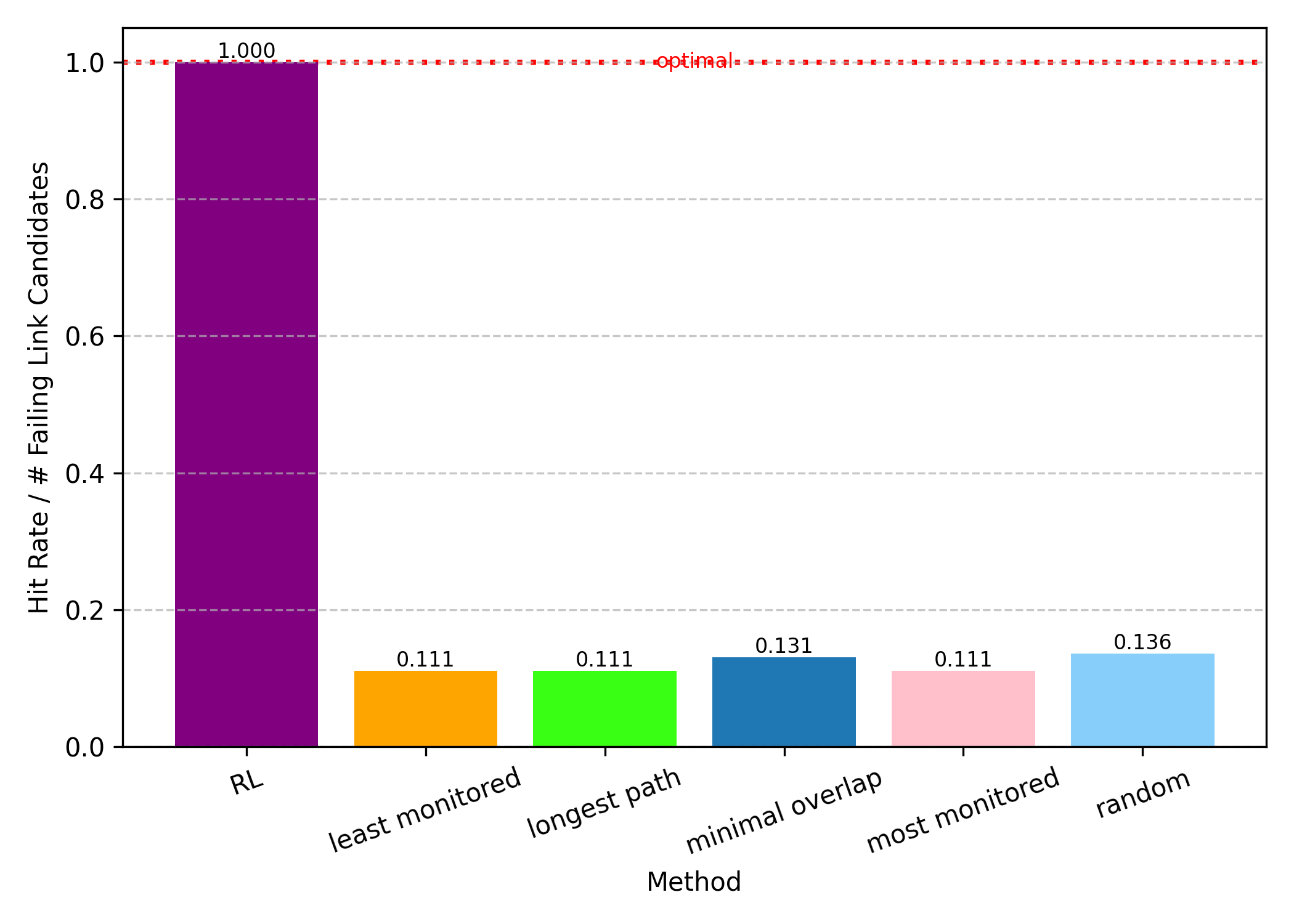}}
    \\
    \caption{Candidate Edges and Efficiency in Topology 2 with OSNR Threshold $t = 16.5$.}
    \label{fig:16.5 OSNR Threshold results} 
\end{figure}

\begin{figure} 
    \centering
    \subfloat[The Size of Candidate Set. (True failing fiber: (dA-v2, dB-v3)).\label{fig:20_0_dA_v2_dB_v3_plot}]{%
        \includegraphics[width=0.5\linewidth]{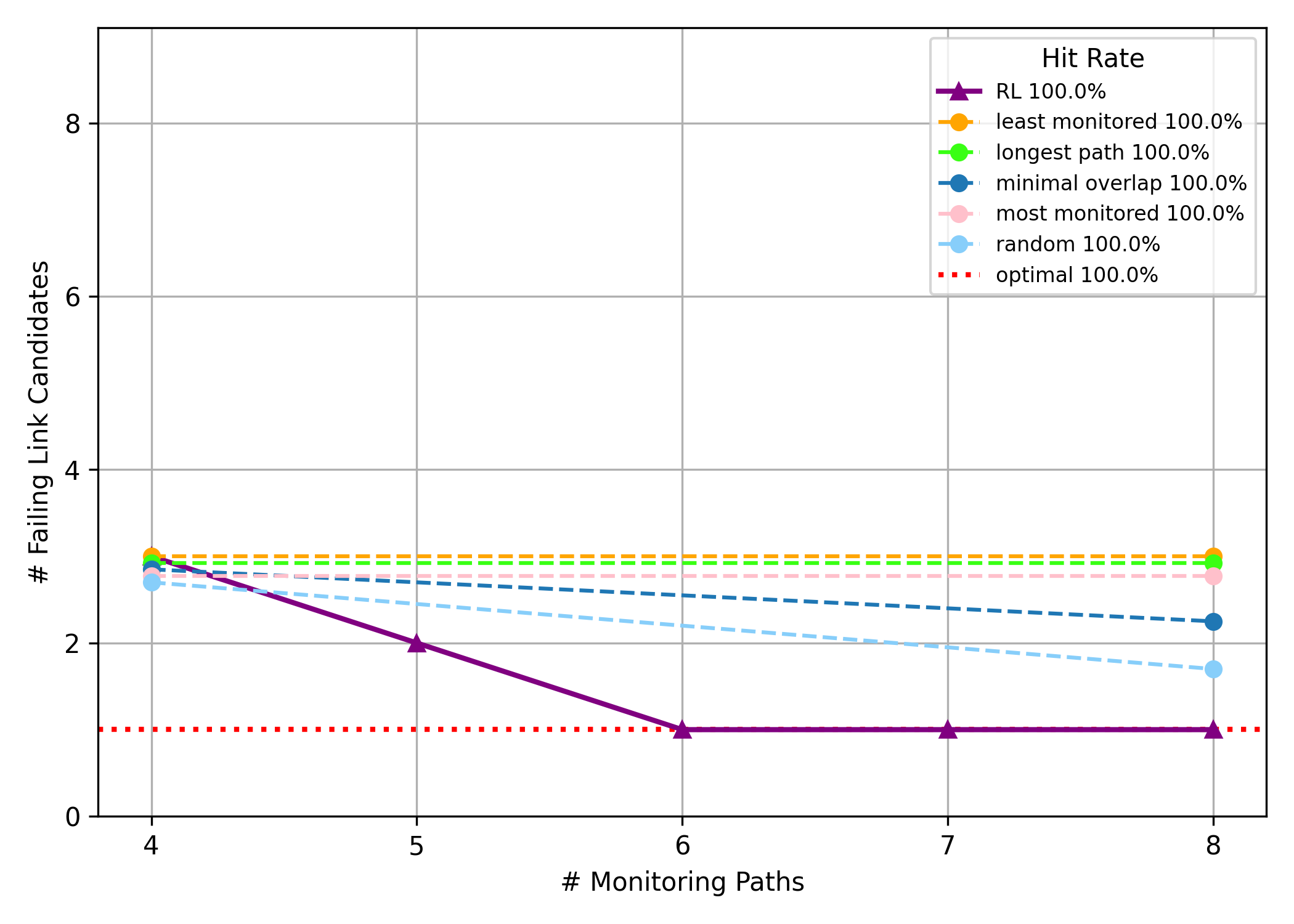}}
    \hfill
    \subfloat[The Efficiency Ratio. (True failing fiber: (dA-v2, dB-v3)).\label{fig:20_0_dA_v2_dB_v3_bar}]{%
        \includegraphics[width=0.5\linewidth]{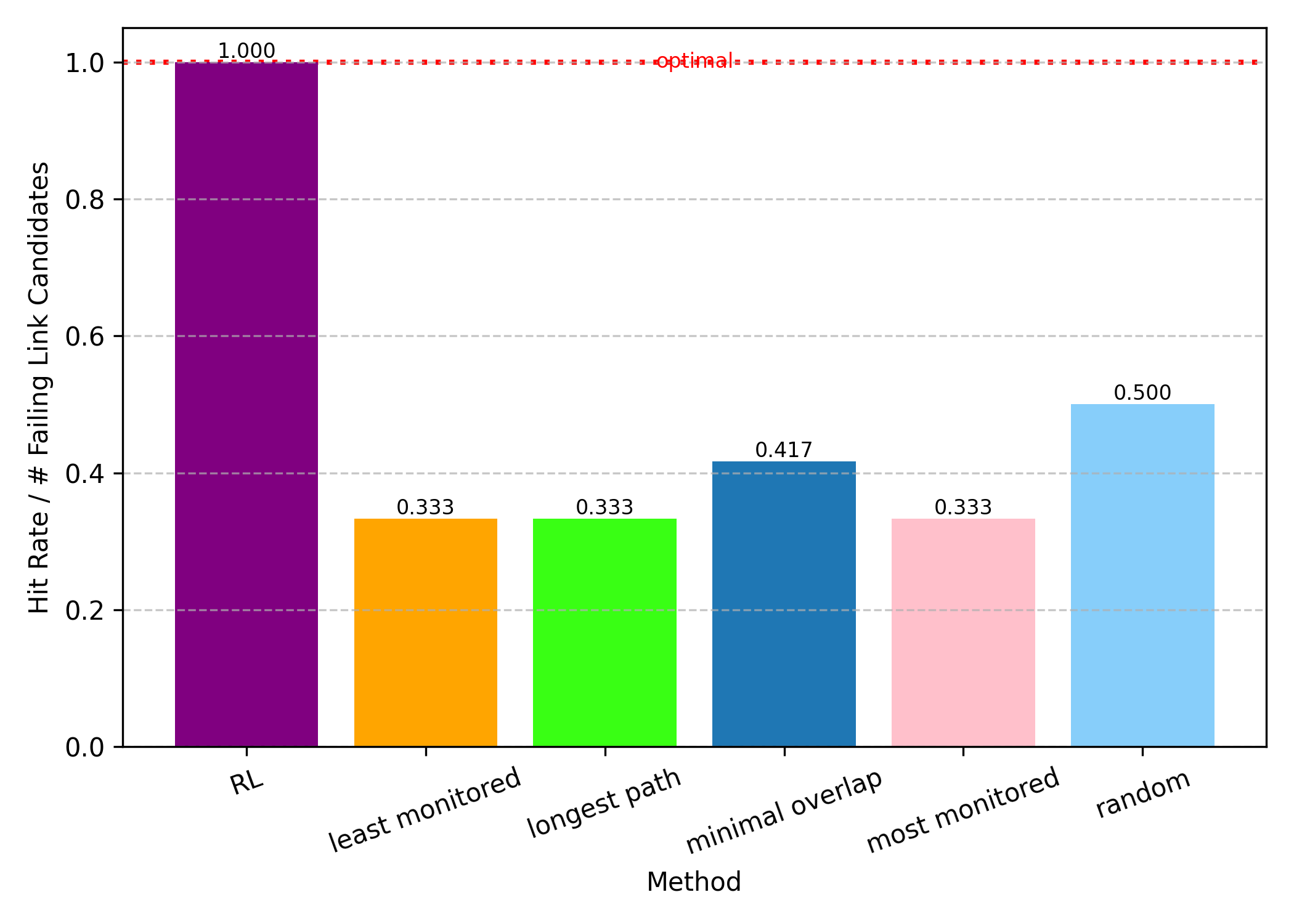}}
    \\
    \subfloat[The Size of Candidate Set. (True failing fiber: (dB-v4,dD-v1)).\label{fig:20_0_dB_v4_dD_v1_plot}]{%
        \includegraphics[width=0.5\linewidth]{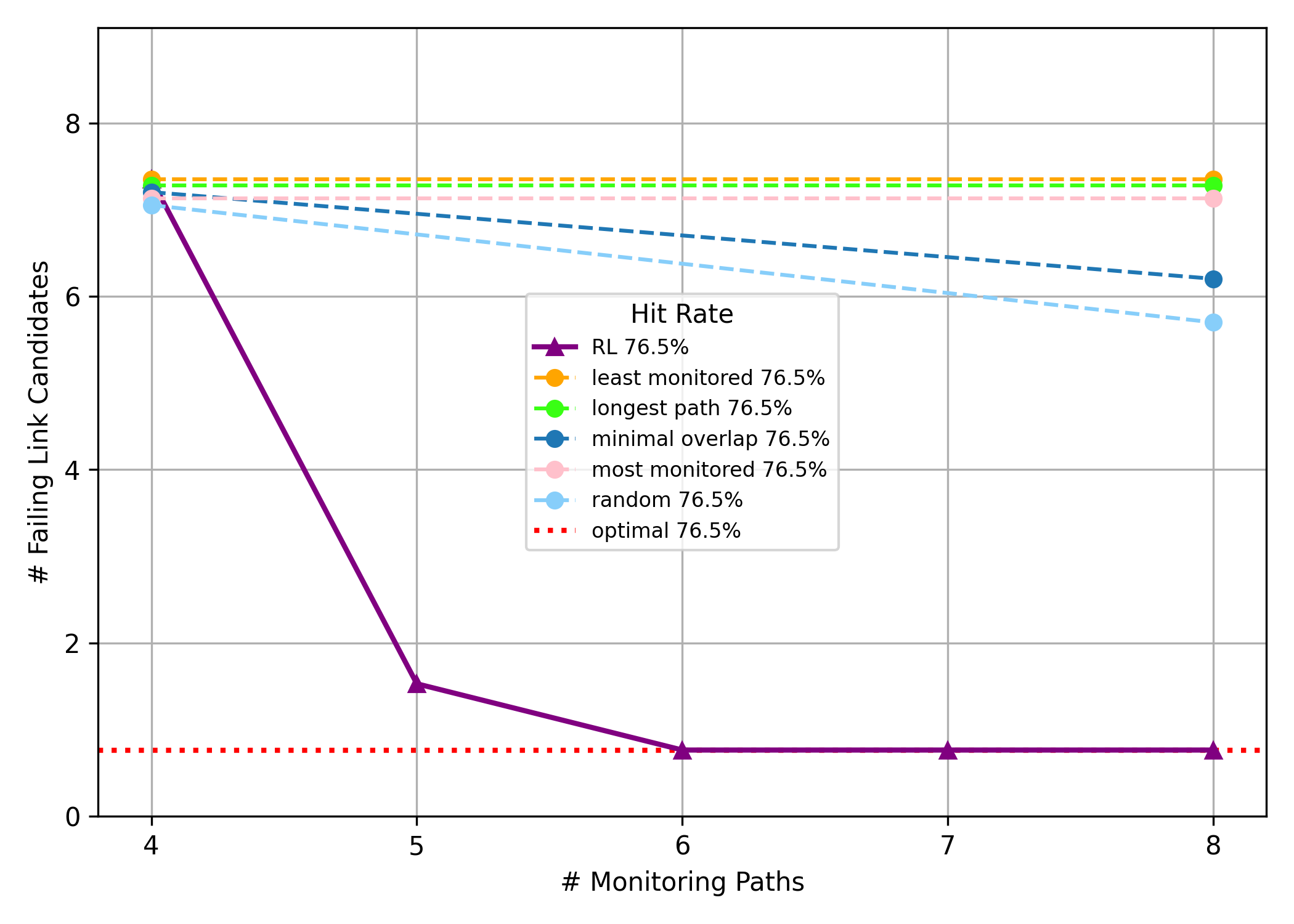}}
    \hfill
    \subfloat[The Efficiency Ratio. (True failing fiber: (dB-v4,dD-v1)).\label{fig:20_0_dB_v4_dD_v1_bar}]{%
        \includegraphics[width=0.5\linewidth]{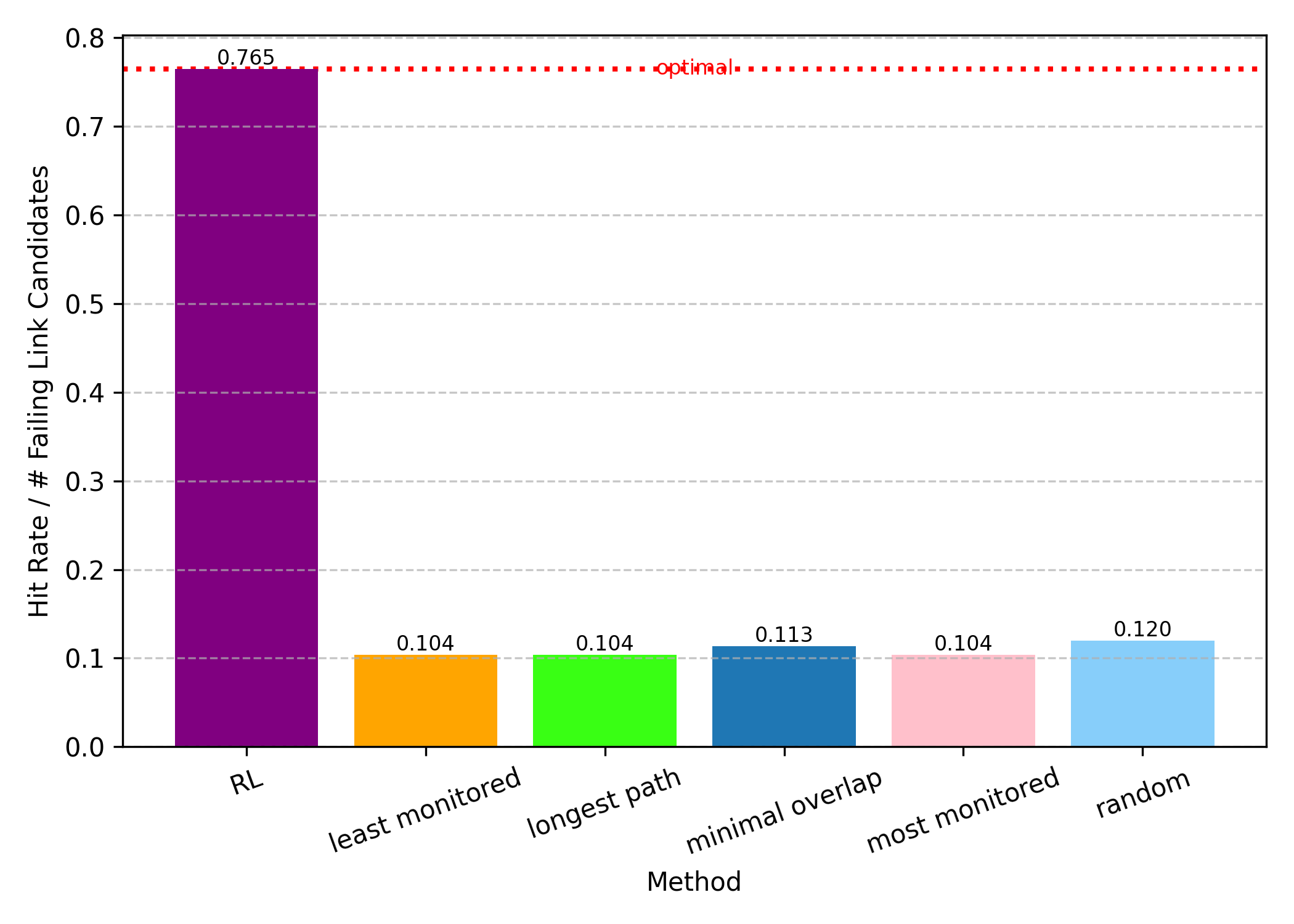}}
    \caption{Candidate Edges and Efficiency in Topology 2 with OSNR Threshold $t = 20.0$.}
    \label{fig:20.0 OSNR Threshold results} 
\end{figure}

\subsection{Progressive Monitoring Path Selection Performance}
We demonstrate the results of the progressive monitoring path selection \textcolor{\revone}{in Topologies 2 and 3} with two different thresholds $t = 16.5$ and $20.0$.
\textcolor{\revone}{
We have tested the RL performance with eight different failure scenarios (randomly selected edges in the broker graph) in Topology 2 and two failure scenarios in Topology 3 in total. We observed two cases in our results. In the trivial cases, the initial monitoring paths already yield perfect or near-perfect localization, and progressive additions of monitoring paths by any method provide no further improvement. In the non-trivial cases, each added path contributes to localization performance in different degrees. In these non-trivial cases, our RL policy consistently achieves the best results. Since the trivial cases produce a flat curve across all methods, the remainder of this section reports only the non-trivial cases, which are the focus of our evaluation.}
For the initial round of the progressive path selection, the same set of monitoring paths was given to all methods by our initial monitoring path selection RL agent that minimizes the LNI. Also, the following experiments assume $K = 4$, i.e., a network operator adds four additional monitoring paths on top of the initial set of monitoring paths.

We report two evaluation metrics: the size of the potentially failing edge set and the hit rate. The size of the potentially failing edge set is the objective function of the progressive path selection problem. However, there is a non-optimal trivial path selection strategy to minimize the objective, which is to eliminate all edges from the candidate set by selecting a less-overlapping signaling path. While this minimizes the candidate set, it is obviously an incorrect solution, since the goal of the problem is to localize the true failing edge. To complement this, we report the \textit{hit rate}, which is defined as the proportion of the resulting potentially failing edge sets in which the true failing edge was included. 
All experiment results below are the average results of 59 simulation runs for both the heuristics and RL approach, using the exact same output files. To ensure meaningful evaluation, runs in which no signaling monitoring paths are present in the initial monitoring path set are excluded from this metric, as no inference regarding the faulty fiber can be made in such cases.
Combined with the objective function, which evaluates how efficiently a method reduces uncertainty, the hit rate assesses how often it arrives at the correct, nontrivial solution. To demonstrate the combined evaluation of the two metrics competing with each other, we report the efficiency score defined as the ratio between the hit rate and the size of the final potentially failing edge set at the final time step $K$. This normalized measure penalizes methods that produce larger candidate sets and the elimination of all edges from the candidate set, thereby favoring approaches that not only include the correct fiber but also minimize ambiguity.


Fig. \ref{fig:16.5 OSNR Threshold results} shows the localization performance of our RL approach and the five baselines \textcolor{\revone}{in Topology 2} with the failure threshold $t = 16.5$ when a fiber (dA-v2, dB-v3) or (dB-v4, dD-v1) is failing. In this case, all methods successfully retained the true failing fiber within the candidate set until the end, which is indicated by the hit rate of 100\%. Focusing on the cardinality of the potentially failing edge set (suspects), our RL approach was able to narrow it down better than the baselines. \textcolor{\revone}{Indeed, our RL approach succeeded in reducing the cardinality to 1, which is the theoretical optimum of perfect localization.} When a fiber (dA-v2, dB-v3) fails, the initial signaling monitoring paths produce approximately three suspects on average. The random and minimum overlap reduced it to around two suspects after adding four paths, while our RL agent produced the perfect localization of identifying the true failing edge at time step $K$ on average. We also observe similar trends when a fiber (dB-v4, dD-v1) fails. In this scenario, the baselines struggled to reduce the number of suspects more than in the previous scenario. This could be because there is a limited number of monitoring path candidates that include the true failing fiber, and it requires more intelligence to find an effective additional path. 


Fig. \ref{fig:20.0 OSNR Threshold results} reports the same localization performance \textcolor{\revone}{in Topology 2} with another threshold $t=20.0$ under the same fiber failure scenarios. Similar to the previous experiment with $t=16.5$, the RL approach consistently outperforms all baselines in both localization performance. The RL agent was able to reduce the number of failing link candidates from an initial set of suspects to a near-minimal set of one suspect within only a few additional monitoring path selections. In contrast, the baselines exhibit little to no reduction in the candidate set size, remaining largely flat across the full range of monitoring paths.

\begin{figure} 
    \centering
    \subfloat[The Size of Candidate Set. (True failing fiber: (dA-v1, dA-v5)).\label{fig:16_5_dA_v1_dA_vC_plot}]{%
        \includegraphics[width=0.5\linewidth]{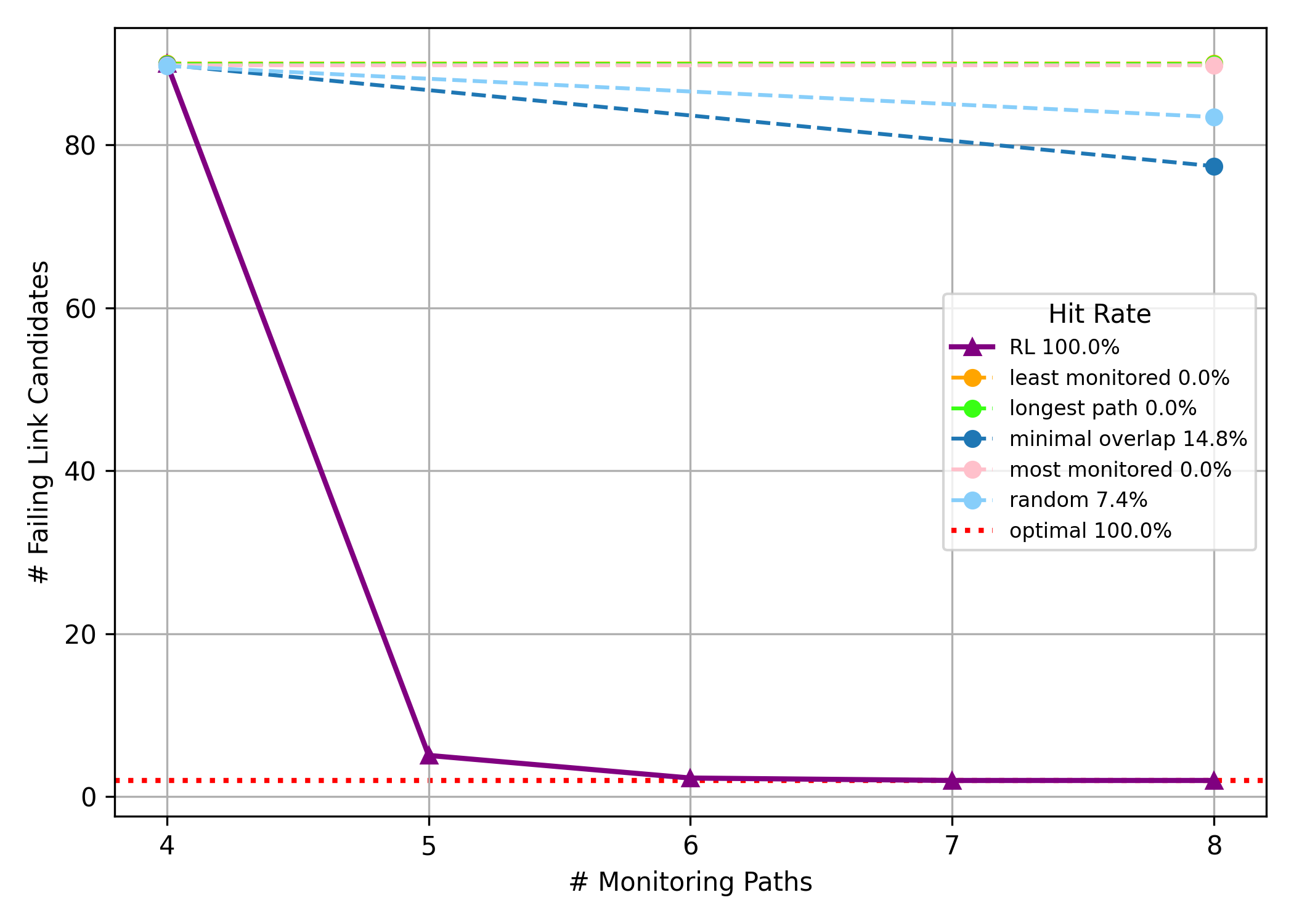}}
    \hfill
    \subfloat[The Efficiency Ratio. (True failing fiber: (dA-v1, dA-v5)).\label{fig:16_5_dA_v1_dA_vC_bar}]{%
        \includegraphics[width=0.5\linewidth]{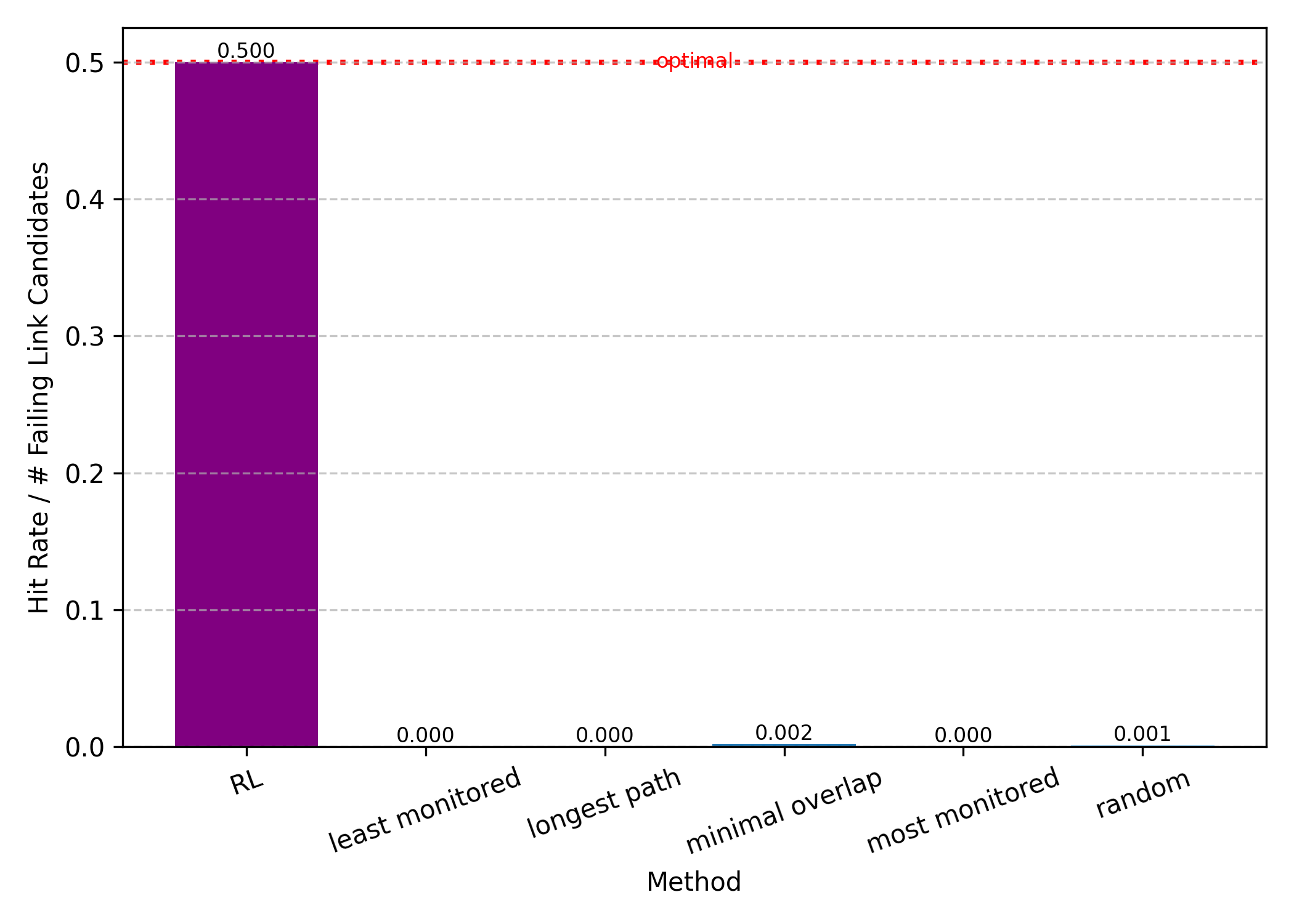}}
    \\
    \subfloat[The Size of Candidate Set. (True failing fiber: (dD-v10, dF-v1)).\label{fig:16_5_dD_v10_dF_v1_plot}]{%
        \includegraphics[width=0.5\linewidth]{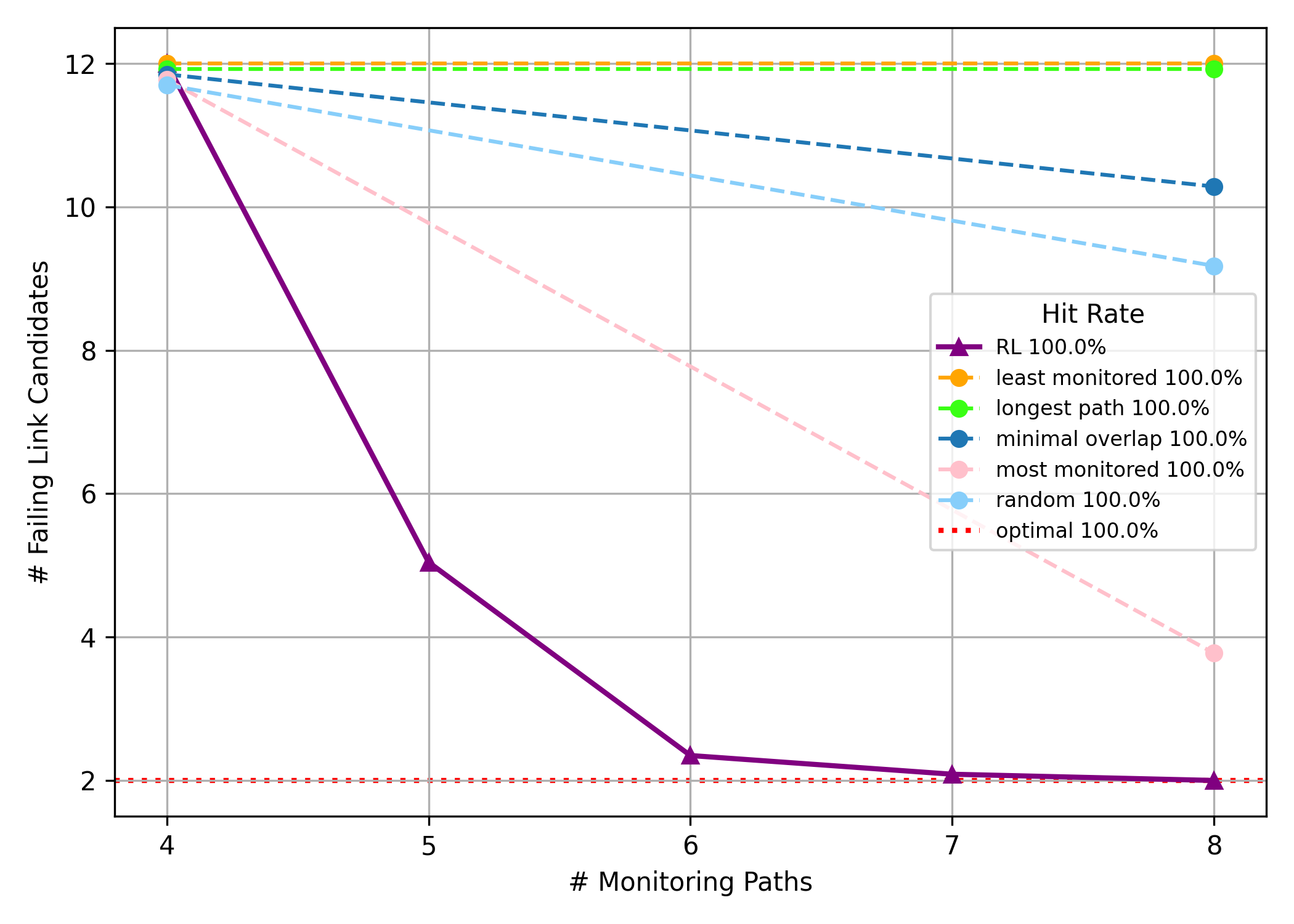}}
    \hfill
    \subfloat[The Efficiency Ratio. (True failing fiber: (dD-v10, dF-v1)).\label{fig:16_5_dD_v10_dF_v1_bar}]{%
        \includegraphics[width=0.5\linewidth]{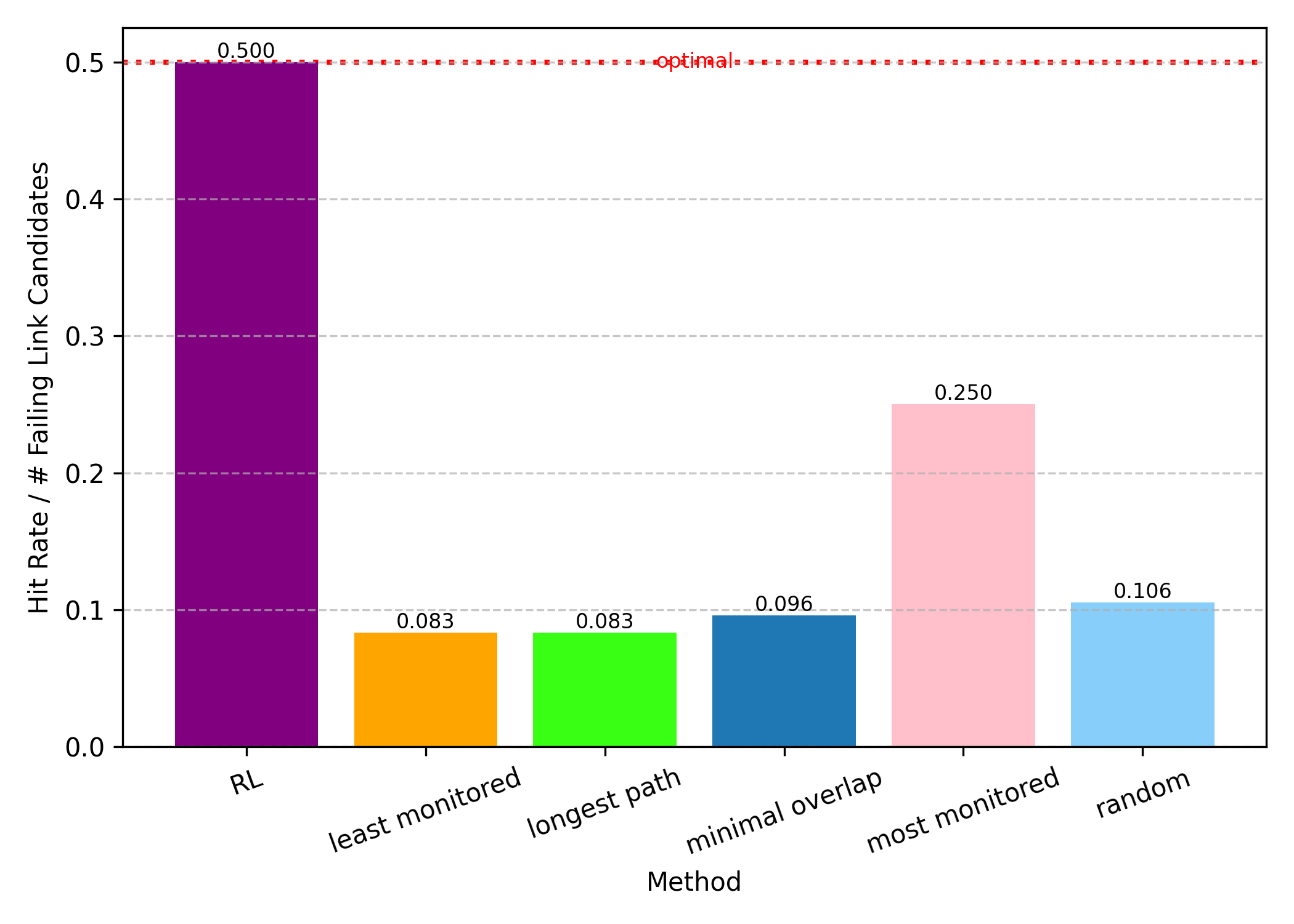}}
    \caption{Candidate Edges and Efficiency in Topology 3 with OSNR Threshold $t = 16.5$.}
    \label{fig:16-5 OSNR Threshold results_topo3} 
\end{figure}

\begin{figure} 
    \centering
    \subfloat[The Size of Candidate Set. (True failing fiber: (dA-v1, dA-vC).\label{fig:20_0_dA_v1_dA_vC_plot}]{%
        \includegraphics[width=0.5\linewidth]{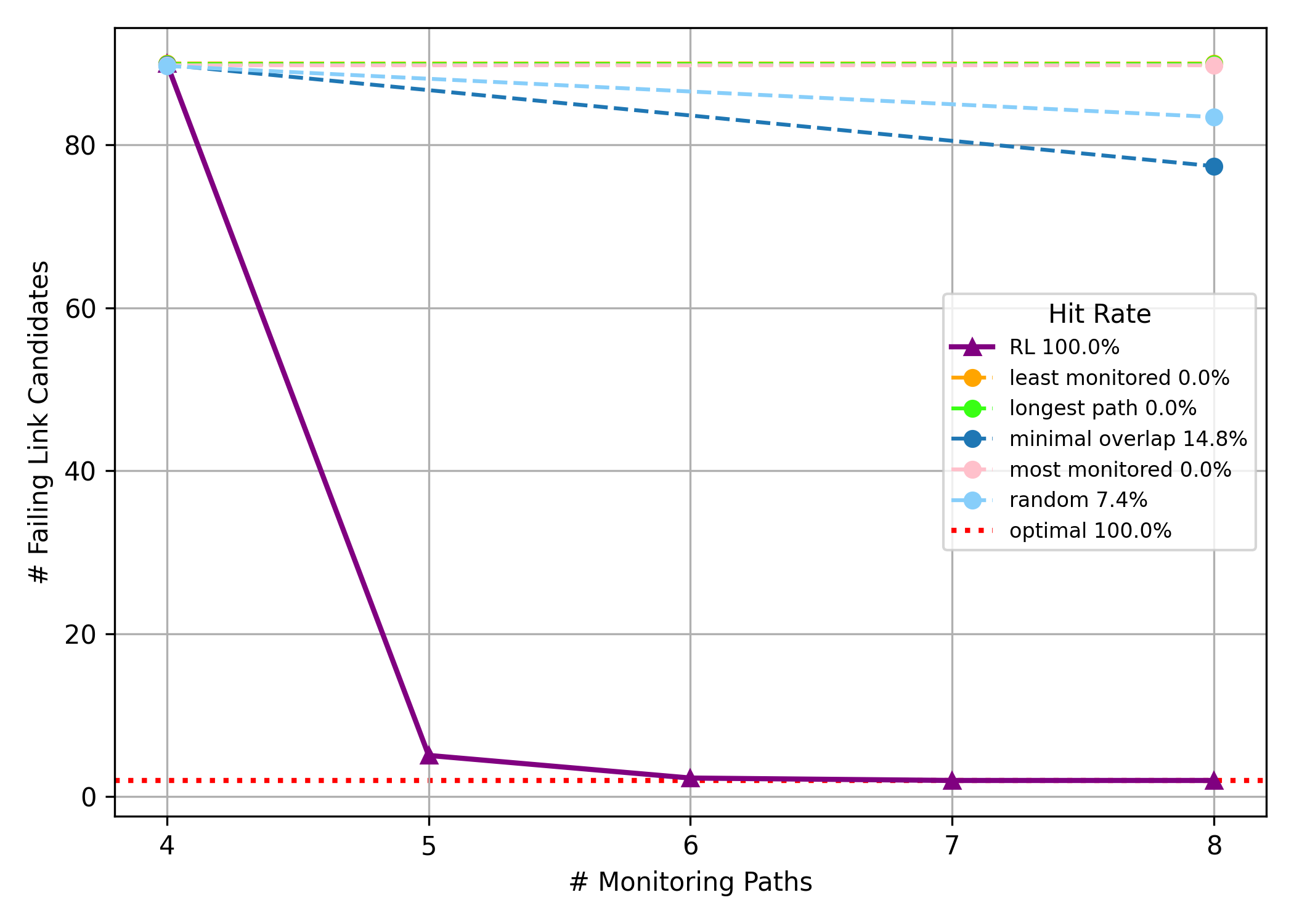}}
    \hfill
    \subfloat[The Efficiency Ratio. (True failing fiber: (dA-v1, dA-vC).\label{fig:20_0_dA_v1_dA_vC_bar}]{%
        \includegraphics[width=0.5\linewidth]{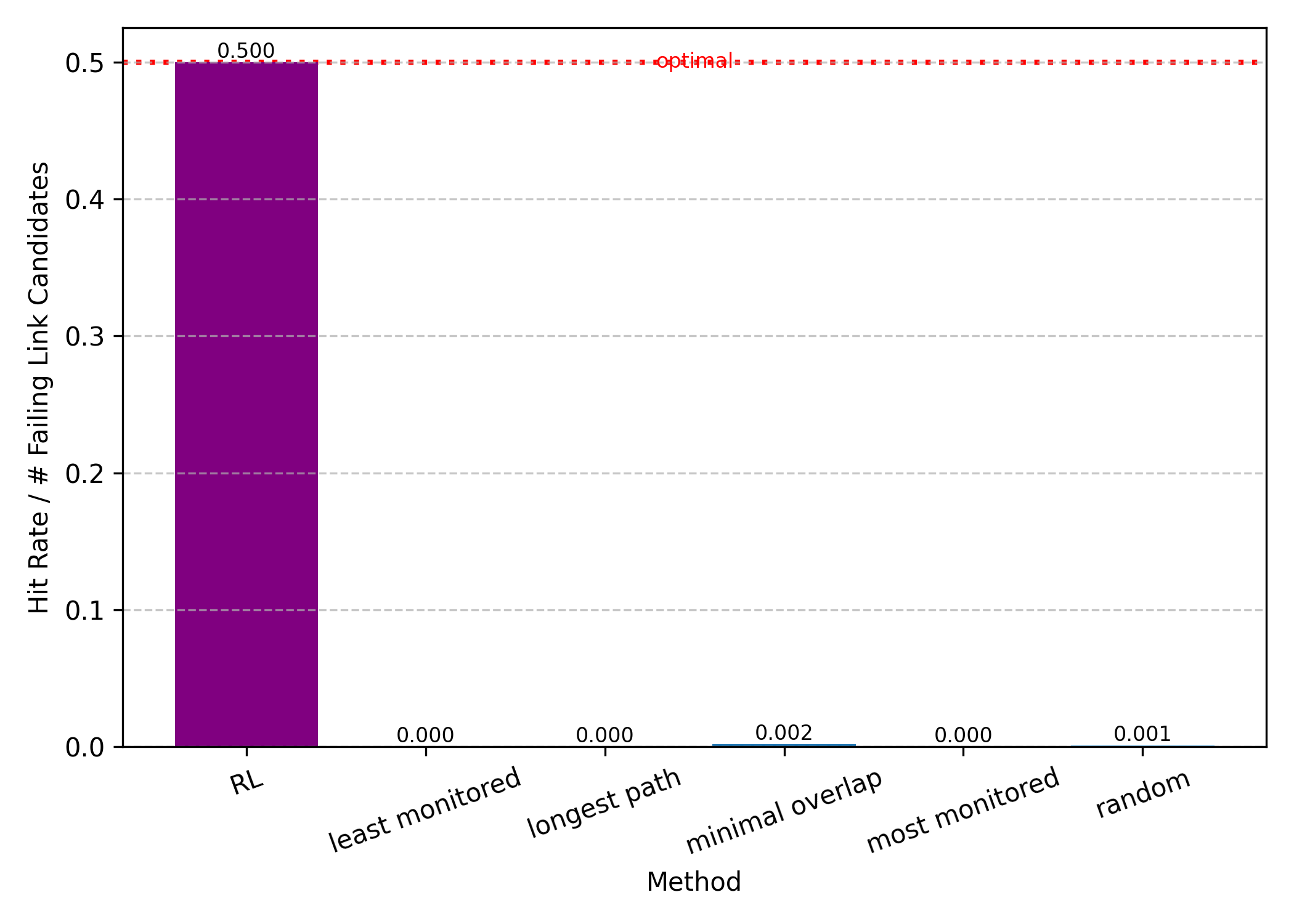}}
    \\
    \subfloat[The Size of Candidate Set. (True failing fiber: (dD-v10, dF-v1).\label{fig:20_0_dD_v10_dF_v1_plot}]{%
        \includegraphics[width=0.5\linewidth]{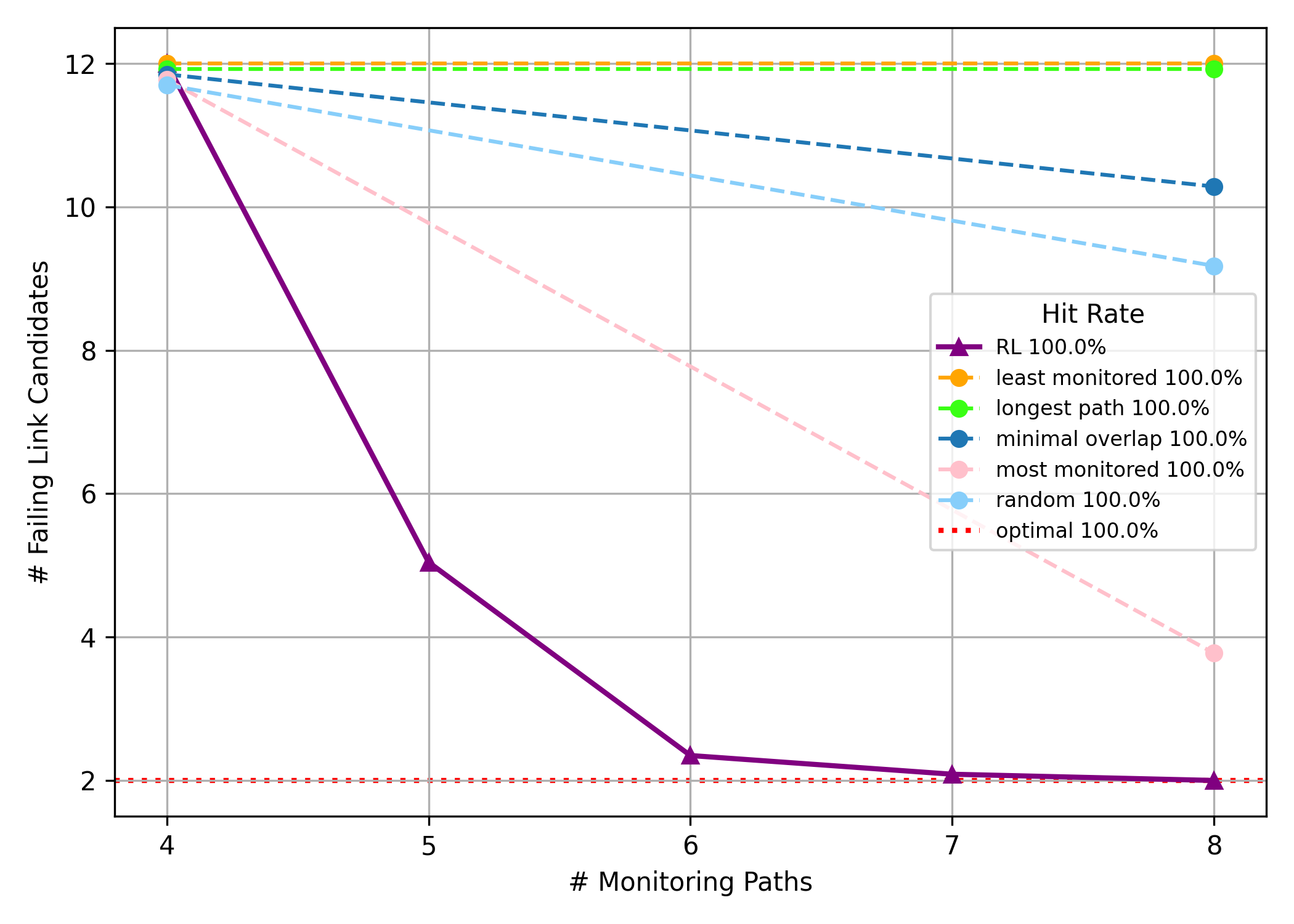}}
    \hfill
    \subfloat[The Efficiency Ratio. (True failing fiber: (dD-v10, dF-v1)).\label{fig:20_0_dD_v10_dF_v1_bar}]{%
        \includegraphics[width=0.5\linewidth]{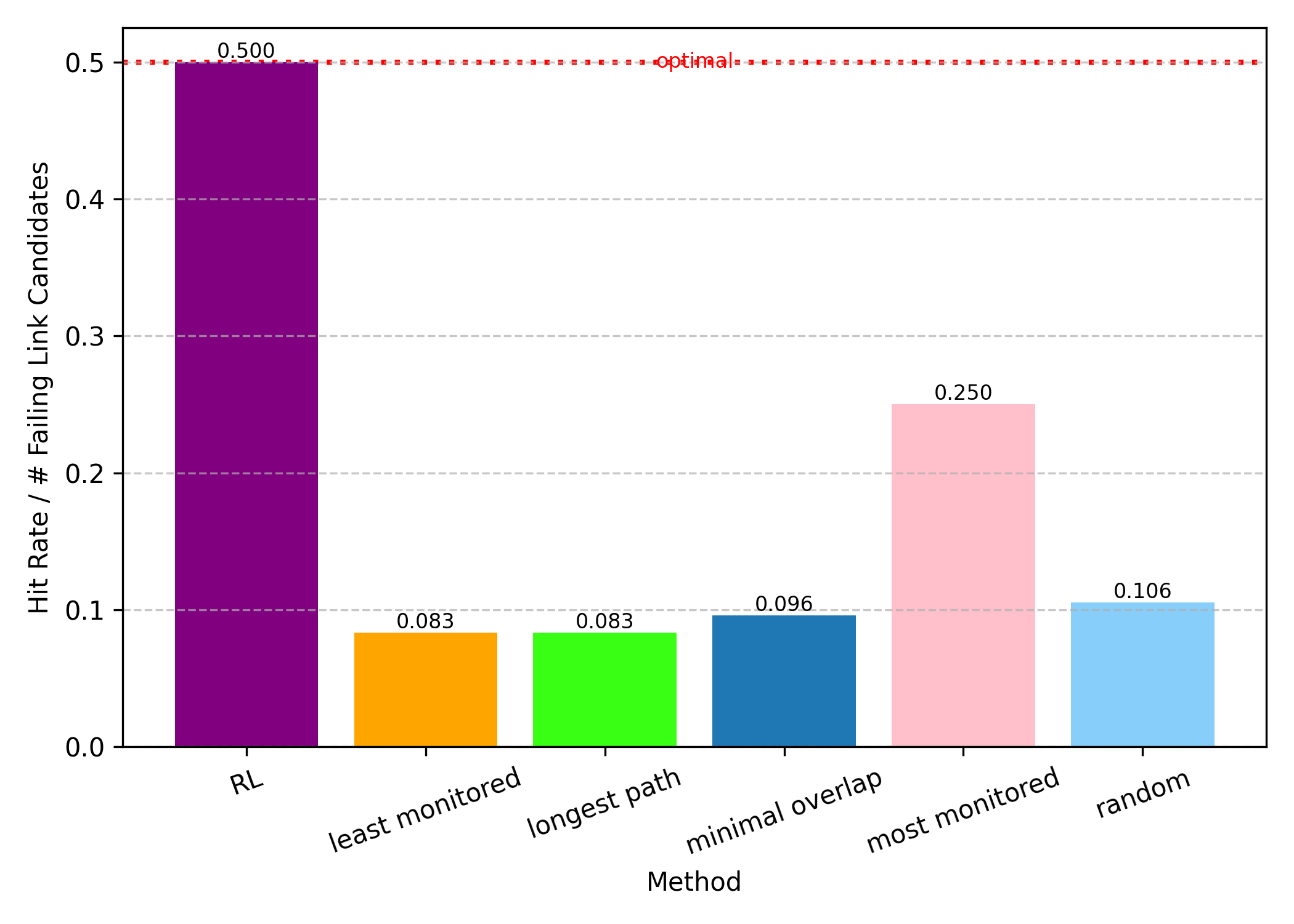}}
    \caption{Candidate Edges and Efficiency in Topology 3 with OSNR Threshold $t = 20.0$.}
    \label{fig:20-0 OSNR Threshold results_topo3} 
\end{figure}

\textcolor{\revone}{
Figs. \ref{fig:16-5 OSNR Threshold results_topo3} and \ref{fig:20-0 OSNR Threshold results_topo3} illustrate the localization performance in Topology 3 with the two threshold values. Since Topology 3 contains more nodes and edges than Topology 2, the number of edges that are suspected to be failing tends to be more than the case in Topology 2. For example, as shown in Fig. \ref{fig:16_5_dA_v1_dA_vC_plot}, 90 edges in the broker graph belong to the suspect set at the initial round with four monitoring paths. This is because none of the monitoring paths is signaling with an OSNR below the threshold value ($t=16.5$). However, as our RL agent strategically adds more monitoring paths, the candidate set drastically decreases in size. Due to the large size of Topology 3, any set of 8 monitoring paths cannot perfectly localize the failing single link. Therefore, the theoretical optimum of the cardinality of the failing link candidates is 2 along with the optimum efficiency ratio of 0.5. Our RL approach was able to achieve these theoretical optima in these scenarios.
}
Overall, by explicitly optimizing to reduce the number of potentially failing edges while preserving consistency with observed failures, the RL agent was able to narrow down the suspects without losing the true failing fiber.


\textcolor{\revone}{We discuss a notable difference between distinct threshold values based on an exemplary scenario where a fiber (dB-v4, dD-v1) fails in Topology 2. With a higher threshold value (e.g., $t=20.0$), more monitoring paths could be flagged as signaling, which increases the risk of misidentification of failing edges.
The probability of this misidentification will be larger than when the threshold is a lower value, like $t=16.5$.
In our results, the hit rates by all methods reduce from 1.0 to 0.765 due to the ambiguity, as shown in Figs. \ref{fig:16_5_dB_v4_dD_v1_bar} and \ref{fig:20_0_dB_v4_dD_v1_bar}.
}

\textcolor{\revone}{In practice, the threshold reflects a network operator's predefined policy on the quality of transmission to be guaranteed and can be derived from historical traces and past monitoring data translated into an appropriate OSNR bound. It is also possible to train multiple agents under different sets of thresholds and select one based on the observed performance. This choice involves an inherent trade-off: a lower threshold is more conservative and certain about path failures, whereas a higher threshold may be able to detect weaker signals of failures. Although not reported in the results section due to space limitations, our preliminary experiments with $t = 18.0$ exhibited no distinctive pattern compared to $t= 20.0$. This also suggests that a fine sweeping granularity is unnecessary and that training only a few agents suffices to compare performance across different combinations of thresholds. Such a selection could itself be automated by a higher-level RL agent that chooses among path-selection agents trained under different thresholds.}

\subsection{Discussions}
\revtwo{
The overall results demonstrate that our RL approach is highly effective at finding where to \textit{spread} monitoring paths for detection and localization of failures at unknown arbitrary locations, and where to \textit{concentrate} monitoring paths for reduction of localization uncertainty after observing initial failure signals. 
The localization capability of a set of monitoring paths is inherently combinatorial. 
As shown in Eq. \eqref{eq:localize}, a failing link is identified by how given signaling paths intersect with each other. Thus, the localization process is set-theoretic and non-additive.
The addition of more paths can worsen the localization capability, even when the monitoring paths before the addition were capable of effectively localizing a failure. In this sense, greedy approaches, including the Least Monitored, have an inherent disadvantage as they optimize local metrics. Furthermore, since the signaling monitoring paths are not known a priori during the initial selection, greedy methods must rely on static topological features. In contrast, RL evaluates the expected long-term performance of each selection of a monitoring path based on past failure experiences. Hence, each selection is made by looking into the probabilistic future possibilities with other dependent path selections, which demonstrates a significant advantage, especially when a larger number of paths are selected in larger topologies.}

\revtwo{
While our RL-based approach demonstrated the capability of effectively selecting monitoring paths, its generalization across network topologies and path configurations remains limited. Because the topologies of optical networks tend to be stable, it is still a possible option for an operator to train a dedicated RL model for its topology. In fact, on a Linux server with an Intel i9 CPU and 64 GB of memory, training converged in approximately 7.3 hours for our largest instance (Topology 3 with 16 monitoring paths). This should be short relative to the timescale on which optical network topologies change. However, the limited number of failure events in practice motivates an improvement in the transferability of a model from one topology to another. Fundamentally, this process requires encoding of topology and path structure to learn abstract characteristics of monitoring paths across diverse topologies. A standard solution includes the insertion of a graph neural network layer into our input layer, which remains as future work.
}

\section{Conclusion}
This paper proposed a deep reinforcement learning-based approach to two types of monitoring path selection problems for E2E optical network slices: the initial and progressive monitoring path selection problems. The former aims at increasing the coverage of the monitoring to improve the detection and localization capability of any failure, while the latter optimizes the selection of monitoring paths to localize a specific failure that has been detected by the initial set of monitoring paths by targeting a suspected topological area. The two path selection problems collectively realize autonomous monitoring, with the goal of maximizing single-link failure localization capability. 
Our method addresses the challenging scenario of link failure localization under limited domain visibility. Experimental results across various network topologies demonstrate that our RL approach consistently outperforms baseline heuristics and even achieves near-optimal localization performance in the initial monitoring path selection problem. The RL agent adapts effectively to different failure situations and learns to select an effective set of monitoring paths for failure localization. As domain autonomy and the service view of domain connectivity become prevalent in multi-domain optical networks, our solution offers a scalable and intelligent framework for E2E performance monitoring. 
 
In future work, we want to extend this work to the emerging disaggregated optical networks with partial visibility, where some domains cooperate with each other by sharing more information, while others stay with the traditional domain model. \textcolor{\revone}{Also, we plan to further improve the RL agent design with graph neural network layers to generalize the learning across different topologies and path configurations without explicit failure thresholds. In particular, the threshold-based definition of signaling paths defined in Eq. \eqref{eq:signalingdef} can be relaxed and extended for more precise failure detection and soft failure scenarios by employing a recurrent neural network-based RL agent that learns directly from historical raw OSNR sequences and translates them into failure alarms. 
}\textcolor{\revone}{
Another natural extension of this work is to generalize the single-link failure model to multiple and correlated failures, such as shared-risk link groups, better reflecting realistic network failure patterns. A major challenge for the extension is the combinatorial growth of a failing edge candidate (suspect) set. With $k$-concurrent failures, the number of failing edge candidate sets scales as $\binom{|E|}{k}$, where $E$ is a set of edges in a broker graph. A path selection problem to disambiguate such concurrent failures needs to reflect the probabilistic nature of signals obtained from the current monitoring paths and interdependencies with additional paths. This could be linked back to the aforementioned future work, where RL learns directly from historical raw OSNR to infer the underlying system state.
}


\begin{backmatter}
\bmsection{Funding}
This work was supported in part by the National Science Foundation under Grant No. CNS-2008856.

\bmsection{Disclosures}
The authors declare no conflicts of interest.





\bmsection{Data availability} Data underlying the results presented in this paper are publicly available at the authors' GitHub repository \cite{adaptive-e2e-monitoring}.

\end{backmatter}

\bibliography{ref}

\end{document}